\documentclass[aip,reprint]{revtex4-1}

\usepackage{graphicx}
\usepackage[font=small]{caption}
\usepackage{subcaption}
\usepackage{multirow}
\usepackage[colorlinks=true, allcolors=blue]{hyperref}
\usepackage{array}
\usepackage[english]{babel}
\usepackage{letltxmacro}

\LetLtxMacro{\ORIGselectlanguage}{\selectlanguage}
\makeatletter
\DeclareRobustCommand{\selectlanguage}[1]{%
  \@ifundefined{alias@\string#1}
    {\ORIGselectlanguage{#1}}
    {\begingroup\edef\x{\endgroup
       \noexpand\ORIGselectlanguage{\@nameuse{alias@#1}}}\x}%
}
\newcommand{\definelanguagealias}[2]{%
  \@namedef{alias@#1}{#2}%
}
\makeatother
\definelanguagealias{en}{english}
\definelanguagealias{eng}{english}

\begin{document}
	

\title{A Transition-edge Sensor-based X-ray Spectrometer for the Study of Highly Charged Ions at the National Institute of Standards and Technology Electron Beam Ion Trap} 

\thanks{This article may be downloaded for personal use only. Any other use requires prior permission of the author and AIP Publishing. This article appeared in Review of Scientific Instruments \textbf{90}, 123107 (2019) and may be found at \url{http://dx.doi.org/10.1063/1.5116717}.}



\author{P.~Szypryt}
\email[]{paul.szypryt@nist.gov}
\affiliation{Quantum Electromagnetics Division, National Institute of Standards and Technology, Boulder, Colorado 80305, USA}

\author{G.~C.~O'Neil}
\affiliation{Quantum Electromagnetics Division, National Institute of Standards and Technology, Boulder, Colorado 80305, USA}

\author{E.~Takacs}
\affiliation{Quantum Measurement Division, National Institute of Standards and Technology, Gaithersburg, Maryland 20899, USA}
\affiliation{Department of Physics \& Astronomy, Clemson University, Clemson, South Carolina 29634, USA}

\author{J.~N.~Tan}
\affiliation{Quantum Measurement Division, National Institute of Standards and Technology, Gaithersburg, Maryland 20899, USA}

\author{S.~W.~Buechele}
\affiliation{Quantum Measurement Division, National Institute of Standards and Technology, Gaithersburg, Maryland 20899, USA}

\author{A.~S.~Naing}
\affiliation{Quantum Measurement Division, National Institute of Standards and Technology, Gaithersburg, Maryland 20899, USA}
\affiliation{Department of Physics \& Astronomy, University of Delaware, Newark, Delaware 19716, USA}

\author{D.~A.~Bennett}
\affiliation{Quantum Electromagnetics Division, National Institute of Standards and Technology, Boulder, Colorado 80305, USA}

\author{W.~B.~Doriese}
\affiliation{Quantum Electromagnetics Division, National Institute of Standards and Technology, Boulder, Colorado 80305, USA}

\author{M.~Durkin}
\affiliation{Quantum Electromagnetics Division, National Institute of Standards and Technology, Boulder, Colorado 80305, USA}
\affiliation{Department of Physics, University of Colorado, Boulder, Colorado 80309, USA}

\author{J.~W.~Fowler}
\affiliation{Quantum Electromagnetics Division, National Institute of Standards and Technology, Boulder, Colorado 80305, USA}
\affiliation{Department of Physics, University of Colorado, Boulder, Colorado 80309, USA}

\author{J.~D.~Gard}
\affiliation{Department of Physics, University of Colorado, Boulder, Colorado 80309, USA}

\author{G.~C.~Hilton}
\affiliation{Quantum Electromagnetics Division, National Institute of Standards and Technology, Boulder, Colorado 80305, USA}

\author{K.~M.~Morgan}
\affiliation{Quantum Electromagnetics Division, National Institute of Standards and Technology, Boulder, Colorado 80305, USA}
\affiliation{Department of Physics, University of Colorado, Boulder, Colorado 80309, USA}

\author{C.~D.~Reintsema}
\affiliation{Quantum Electromagnetics Division, National Institute of Standards and Technology, Boulder, Colorado 80305, USA}

\author{D.~R.~Schmidt}
\affiliation{Quantum Electromagnetics Division, National Institute of Standards and Technology, Boulder, Colorado 80305, USA}

\author{D.~S.~Swetz}
\affiliation{Quantum Electromagnetics Division, National Institute of Standards and Technology, Boulder, Colorado 80305, USA}

\author{J.~N.~Ullom}
\affiliation{Quantum Electromagnetics Division, National Institute of Standards and Technology, Boulder, Colorado 80305, USA}
\affiliation{Department of Physics, University of Colorado, Boulder, Colorado 80309, USA}

\author{Yu.~Ralchenko}
\affiliation{Quantum Measurement Division, National Institute of Standards and Technology, Gaithersburg, Maryland 20899, USA}


\begin{abstract}
We report on the design, commissioning, and initial measurements of a Transition-edge Sensor (TES) x-ray spectrometer for the Electron Beam Ion Trap (EBIT) at the National Institute of Standards and Technology (NIST). Over the past few decades, the NIST EBIT has produced numerous studies of highly charged ions in diverse fields such as atomic physics, plasma spectroscopy, and laboratory astrophysics. The newly commissioned NIST EBIT TES Spectrometer (NETS) improves the measurement capabilities of the EBIT through a combination of high x-ray collection efficiency and resolving power. NETS utilizes 192 individual TES x-ray microcalorimeters (166/192 yield) to improve upon the collection area by a factor of $\sim$30 over the 4-pixel neutron transmutation doped germanium-based microcalorimeter spectrometer previously used at the NIST EBIT. The NETS microcalorimeters are optimized for the x-ray energies from roughly 500~eV to 8,000~eV and achieve an energy resolution of 3.7~eV to 5.0~eV over this range, a more modest ($<2 \times$) improvement over the previous microcalorimeters. Beyond this energy range NETS can operate with various trade-offs, the most significant of which are reduced efficiency at lower energies and being limited to a subset of the pixels at higher energies. As an initial demonstration of the capabilities of NETS, we measured transitions in He-like and H-like O, Ne, and Ar as well as Ni-like W. We detail the energy calibration and data analysis techniques used to transform detector counts into x-ray spectra, a process that will be the basis for analyzing future data.
\end{abstract}

\pacs{}

\maketitle 

\section{Introduction}
\label{sec:Introduction}

The Electron Beam Ion Trap (EBIT\cite{levine_electron_1988, vogel_design_1990, becker_electron_2010}) is a powerful laboratory instrument designed to create, store, and excite highly charged ions that may be studied through their characteristic emission spectra. In the EBIT, electrons are generated by an electron gun, and a series of electrodes along the central axis of the machine accelerates them to high kinetic energies. A set of drift tubes along the beam path sets up a potential well that is used to trap the ions axially. High field magnets, historically liquid helium cooled superconducting magnets, are used primarily to compress the electron beam, which increases the strength of the radial confinement that the negatively charged electron beam provides to the positively charged ion cloud. At the end of the beam path, a collector, which is also typically actively cooled, dissipates the remaining energy of the intense beam.  Many different mechanisms can be used to inject ions into the electron beam of the EBIT. A ballistic gas injection system can be used for studying charged ions of gaseous neutral atoms such as nitrogen, neon, or argon. For metal ions, a more complicated tool must be used, such as a Metal Vapor Vacuum Arc (MeVVA\cite{brown_metal_1992, holland_low_2005}) external ion source. The same electron beam used to trap and ionize the ions also collisionally excites them.

The first EBIT was developed at the Lawrence Livermore National Laboratory (LLNL) in 1985\cite{levine_electron_1988} and was soon after used to take atomic spectroscopy measurements of highly charged barium\cite{marrs_measurement_1988}. This promising new instrument led a handful of groups around the world to build their own EBIT systems. One such system was built at the National Institute of Standards and Technology (NIST) in 1993 and quickly began producing results\cite{gillaspy_first_1997}. Since its inception, the NIST EBIT has been used to take measurements across a variety of fields, including fundamental atomic physics\cite{ralchenko_accurate_2006, gillaspy_precision_2014, payne_helium-like_2014}, spectroscopy of highly-charged ions for fusion and lithography\cite{draganic_euv_2011, ralchenko_spectroscopy_2011, kilbane_euv_2012}, laboratory astrophysics\cite{silver_laboratory_2000, takacs_astrophysics_2003, tan_electron_2005, chen_3c/3d_2006, gillaspy_fe_2011}, and nuclear physics\cite{silwal_measuring_2018}.

Low-temperature x-ray microcalorimeters that combine high resolving power comparable to wavelength dispersive spectrometers with high broadband x-ray collection efficiency comparable to semiconductor based solid-state detectors are particularly well suited for measurements with an EBIT\cite{betancourt-martinez_transition-edge_2014, doriese_practical_2017}. First, an EBIT's tunable energy electron beam is capable of exciting a wide range of charge states. The spectral features from these charge states can be found at energies up to 10s--100s of keV, depending on the highest charge state the particular EBIT can achieve. The broad energy range of an x-ray microcalorimeter allows for the study of a wide array of highly charged ions with a single spectrometer. Furthermore, the exceptional energy resolution of an x-ray microcalorimeter also allows for high precision studies of line emission structures and allows us to resolve previously overlapping features. Finally, the single photon detection nature of a microcalorimeter allows for time-resolved studies with the EBIT. Measurements with microcalorimeters have been highly successful at multiple EBIT or other highly charged ion facilities (see Ref.~\citenum{porter_astro-e2_2004, porter_xrs_2008, beiersdorfer_brief_2008, porter_performance_2008, shen_status_2011, kraft-bermuth_microcalorimeters_2018} and the references therein). Generally, 10s of few 100~$\mu$m scale microcalorimeters are operated within a spectrometer and energy resolutions of $\sim$5~eV at 6~keV are typically observed. At the NIST EBIT, we have utilized an array of 4 x-ray microcalorimeters based on neutron transmutation doped germanium (NTD-Ge\cite{larrabee_ntd_1984}) thermistors and tin absorbers developed by the Harvard Smithsonian Astrophysical Observatory (SAO)\cite{bandler_ntd-ge-based_2000, tan_electron_2005, chen_3c/3d_2006, ralchenko_accurate_2006, gillaspy_fe_2011}. These NTD-Ge microcalorimeters individually have an active area of 350~$\mu$m~$\times$~350~$\mu$m and show a resolution of 4.5~eV at 6~keV during operation with the EBIT\cite{ tan_electron_2005}.

Advances in low temperature detector technology have recently increased the capabilities of spectrometers based on x-ray microcalorimeters, both in terms of total array active area and per-pixel performance. One such technology that is the basis for the new spectrometer at the NIST EBIT is the Transition-edge Sensor (TES\cite{andrews_attenuated_1942, irwin_x-ray_1996}). TES-based x-ray microcalorimeter spectrometers with 100s of pixels have been used for applications in x-ray astronomy\cite{holland_transition_1999, den_herder_x-ray_2012}, beamline science\cite{heates_collaboration_first_2016, doriese_practical_2017, lee_transition-edge_2019}, and tabletop spectroscopy\cite{miaja-avila_ultrafast_2016, oneil_ultrafast_2017}, among other fields. TES microcalorimeters regularly achieve a resolving power of better than 1 part in a 1000 for x-rays in the few keV range\cite{smith_small_2012, bandler_advances_2013, lee_fine_2015, uhlig_high-resolution_2015, ullom_review_2015, doriese_practical_2017} and can measure the arrival time of an x-ray with $\sim$1~$\mu$s timing resolution\cite{heates_collaboration_first_2016}. Other low temperature x-ray microcalorimeter technologies based on Microwave Kinetic Inductance Detectors (MKIDs\cite{day_broadband_2003, ulbricht_highly_2015}) and Metallic Magnetic Calorimeters (MMCs\cite{fleischmann_metallic_2005, fleischmann_physics_2009}) are developing rapidly as well. At the time of this writing, TES-based microcalorimeters have achieved better resolution compared to MKID-based microcalorimeters, and TES multiplexing techniques are more mature compared to those of MMCs. In addition to the NIST EBIT, some of these more recent advances in low temperature detector technology are being utilized in facilities for other highly charged ion experiments. The EBIT at Lawrence Livermore National Laboratory (LLNL) is in the process of adding a TES x-ray spectrometer to their detector suite\cite{betancourt-martinez_transition-edge_2014}. Also, MMCs with high dynamic range have recently been used to study highly charged ions at the Experimental Storage Ring (ESR) at the GSI Helmholtz Centre for Heavy Ion Research\cite{kraft-bermuth_microcalorimeters_2018}. 

Here we detail the design, commissioning, and first-light measurements of the NIST EBIT TES Spectrometer (NETS). NETS has a factor of $>30$ increase in the active area of the array while also exhibiting a modest ($<2 \times$) improvement in energy resolution when compared to the to NTD-Ge microcalorimeter spectrometer previously used at the NIST EBIT. In Sec.~\ref{sec:TES_Spectrometer}, we explain the physics behind TES-based x-ray microcalorimeters and the specifics of the spectrometer developed for the NIST EBIT. We move on to show how the spectrometer was integrated with the rest of the EBIT in Sec.~\ref{sec:Integration}. We then discuss first-light measurements done with NETS (Sec.~\ref{sec:Measurements}) and techniques used to analyze these data (Sec.~\ref{sec:Reduction}). Finally, in Sec.~\ref{sec:Results} we discuss spectra acquired during the first-light run including the achieved count rates, energy resolution, line center accuracy, and time resolution.

\section{TES Spectrometer}
\label{sec:TES_Spectrometer}

\subsection{TES Microcalorimeter Principles of Operation}

\begin{figure}
\centering
\begin{subfigure}{\linewidth}
    \includegraphics[width=0.9\linewidth]{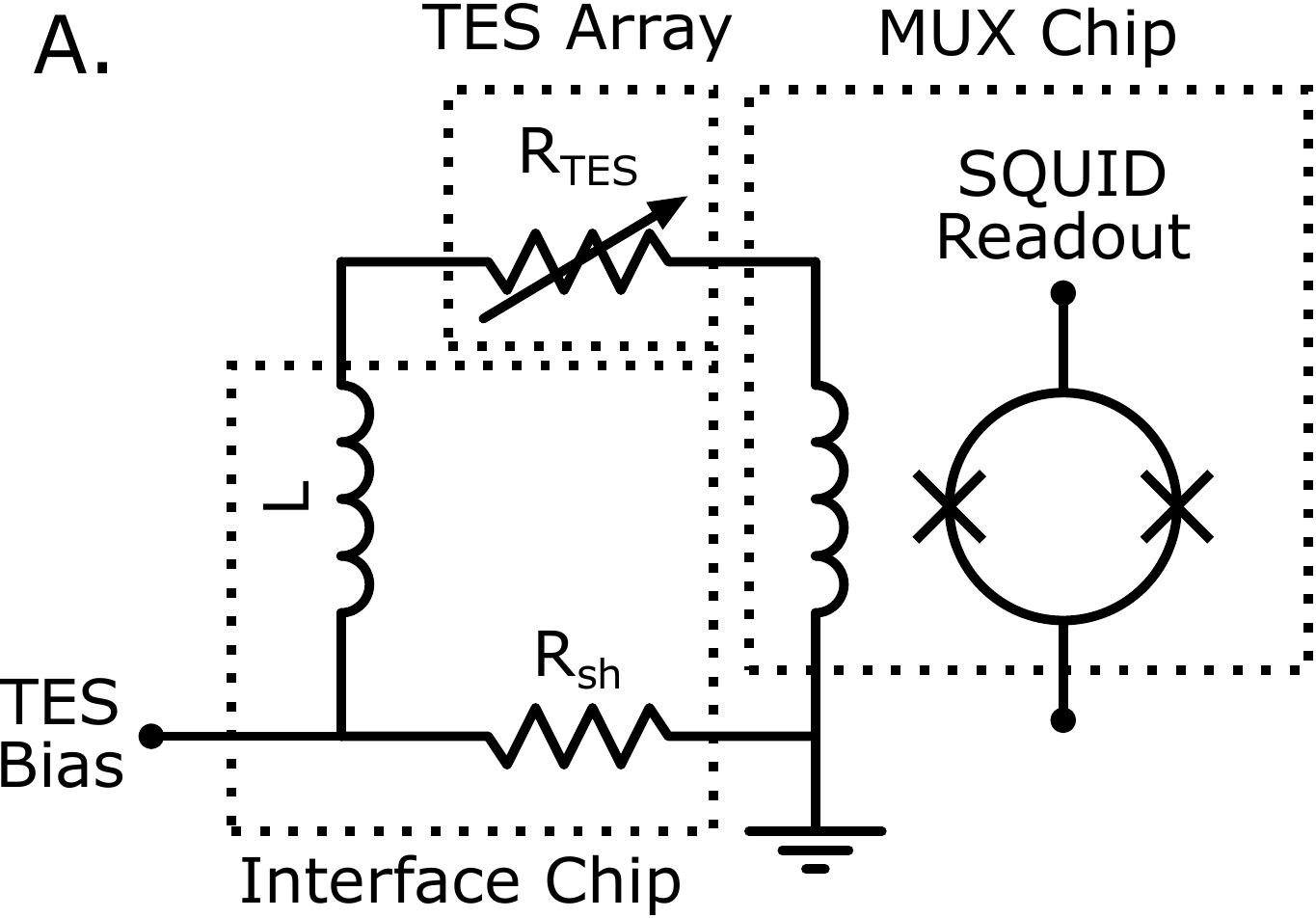}
\end{subfigure}
\par\medskip
\begin{subfigure}{\linewidth}
    \includegraphics[width=1.0\linewidth]{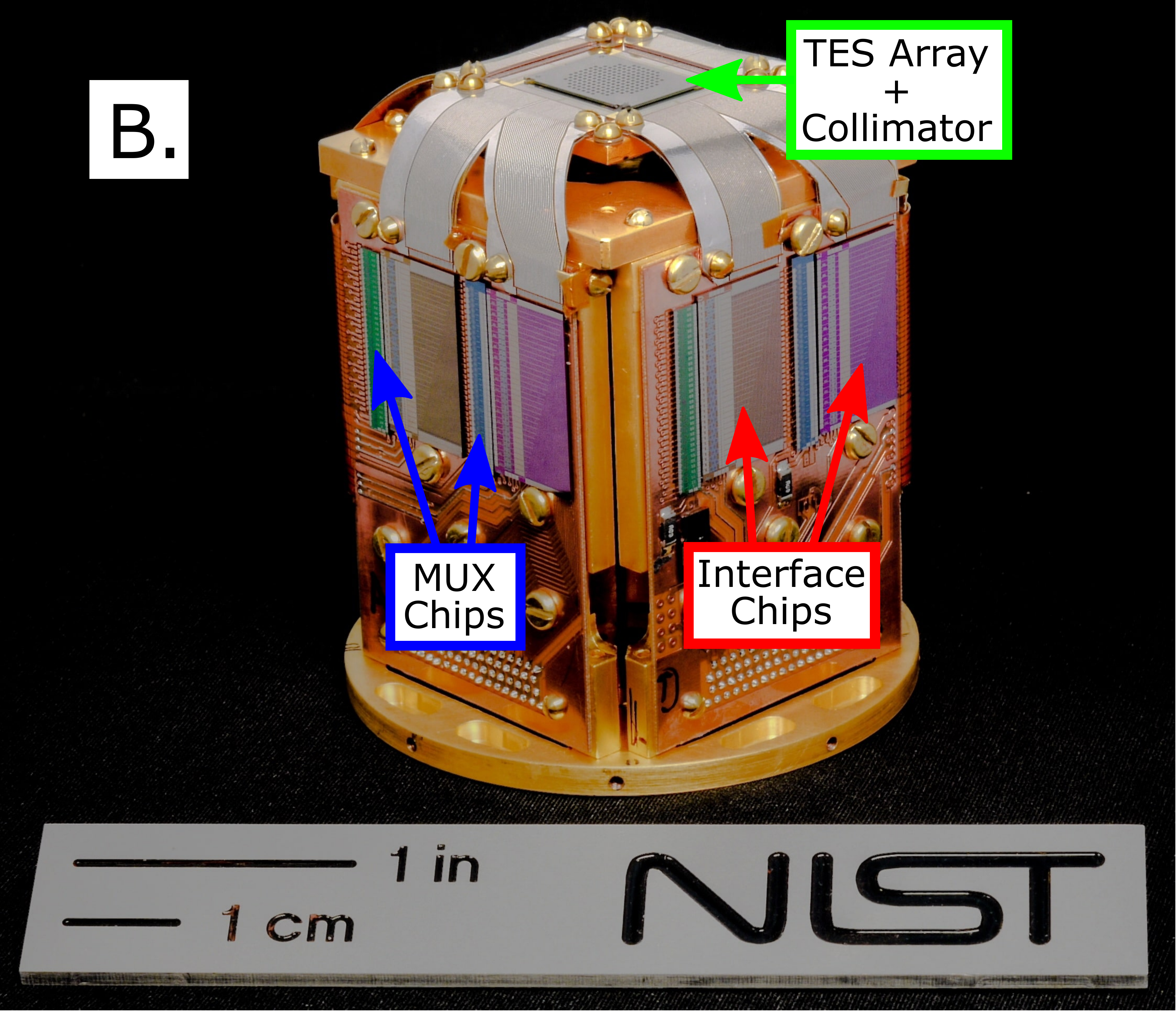}
\end{subfigure}
\caption{\label{fig:TES_Operation} (A)~TES circuit diagram. The TES is voltage biased into its superconducting transition region by sourcing current through a 380~$\mu \Omega$ shunt resistor, $R_{sh}$. When an x-ray event occurs, the TES temperature rises proportionally to the x-ray energy, pushing the TES higher into its transition and raising its resistance (represented by the variable resistor, $R_{TES}$, in the diagram). This results in a pulse of decreased current. An inductor, L, is used to adjust the rise time of this pulse. Each TES is inductively coupled to and read out using a first stage SQUID, and a SQUID-based multiplexing scheme is used to read out the entire array of TESs. (B)~Detector and cold stage readout assembly, including the TES array with collimator and 8 sets of multiplexer (MUX) and interface chips for the 8 independent readout columns. The dashed boxes in the top panel correspond to the different types of physical chips labeled in the bottom panel.}
\end{figure}

The microcalorimeters in NETS use TESs as the sensing elements, weakly coupled to a thermal bath. The devices are cooled to low temperatures, often below 100~mK, to minimize thermal noise and maximize detector sensitivity. Generally, temperatures below the critical temperature ($T_C$) of the TESs are used, and the TESs are then biased into their superconducting-to-normal transition. Within this narrow region, a TES has a high temperature coefficient of resistance (e.g., NIST TES designs typically have $\partial \log R / \partial \log T$ values of $\lesssim$100 at their bias point\cite{doriese_practical_2017}), allowing for a sensitive measurement of the temperature change that occurs during a photon absorption event. A TES used in x-ray microcalorimeters is typically voltage biased into its transition, providing negative electrothermal feedback (power in the TES decreases as absorbed photons raise its resistance). When the TES is voltage-biased, the measured signal from an absorbed photon is a pulse of decreased TES current. TES operation is outlined in Fig.~\ref{fig:TES_Operation}. The physics and operation of TES-based x-ray microcalorimeters are explained in much greater detail in a number of publications\cite{irwin_transition-edge_2005, ullom_review_2015, doriese_practical_2017}.

In order to maximize x-ray collection efficiency, the TES is coupled to a thicker absorber made with a high-Z material, necessary for stopping x-rays with energies above a few hundred eV while maintaining the TES's superconducting and thermodynamic properties. The thermodynamic properties of the absorber, such as the heat capacity, can then be tuned separately from that of the TES, allowing for a simpler optimization of the devices to target a specific energy range. The device is typically fabricated on top of a membrane in order to control thermal conductance (speed of current pulses) and minimize the amount of energy going into phonons escaping into the substrate instead of measurable energy in the sensor. This is necessary for maintaining a high signal-to-noise ratio (SNR) in the pulses and therefore high resolving power.

The readout of TES microcalorimeter arrays is typically accomplished with Superconducting Quantum Intereference Devices (SQUIDs\cite{clarke_squid_2004}). Here, SQUIDs are used in an ammeter mode to measure and amplify the current pulses. In cryogenic detectors, it is often necessary to use a multiplexing scheme to reduce the complexity in wiring and the thermal load inside the cryostat, and SQUIDs are also used for this purpose. A variety of SQUID-based multiplexing architectures\cite{kiviranta_squid-based_2002} can be used, but the most mature readout method is time division multiplexing (TDM\cite{de_korte_time-division_2003, doriese_developments_2016}), which involves sequentially addressing the readout SQUIDs for a group of detectors. NIST has routinely used this method to multiplex $\sim$30 detectors (rows), with 8 parallel readout channels (columns), enabling operation of large arrays of TES microcalorimeters\cite{doriese_practical_2017, bennett_high_2012}.

\subsection{Detector Array}
\label{subsec:DetectorArray}

\begin{figure}
\includegraphics[width=0.85\linewidth]{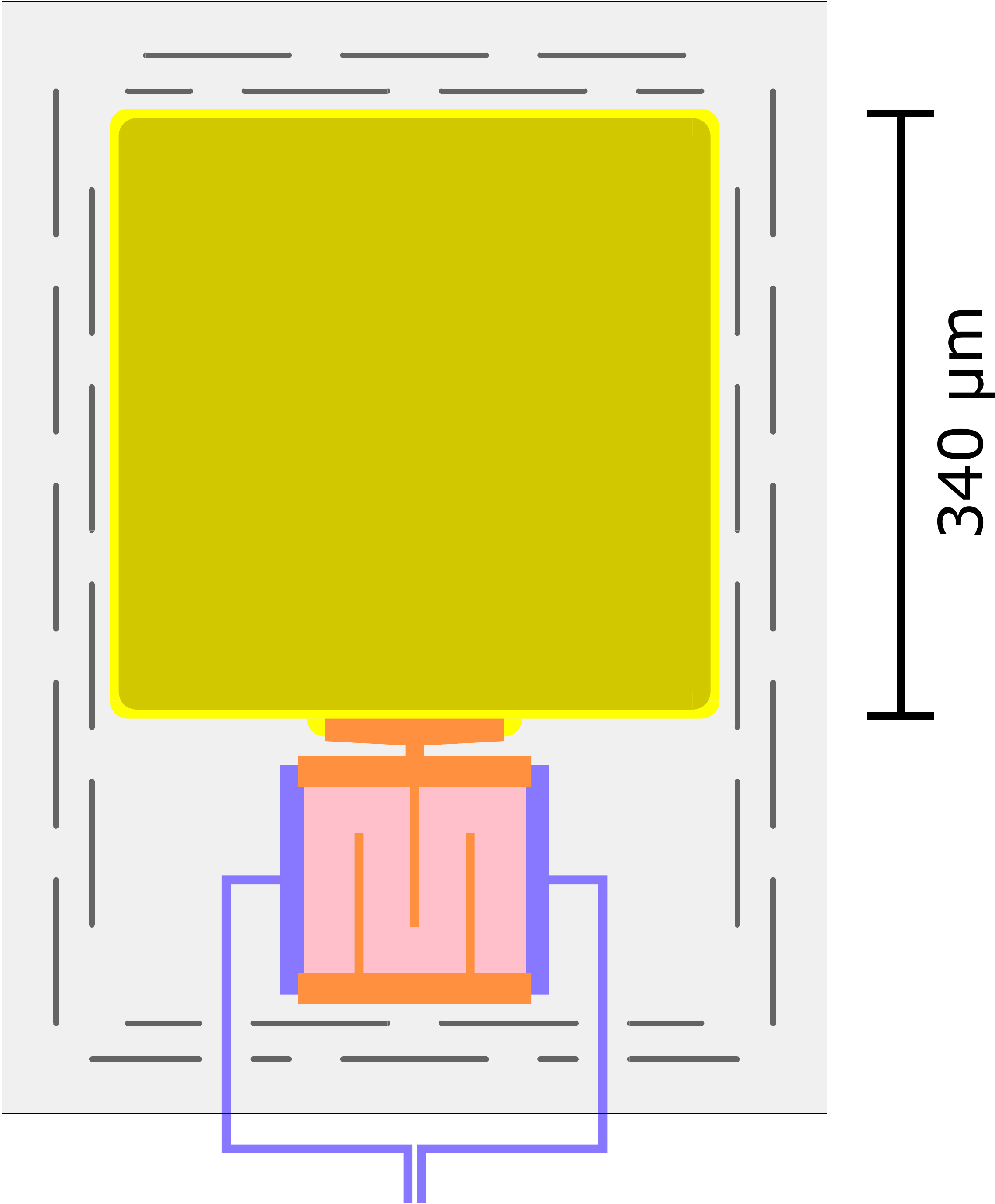}
\caption{\label{fig:TES_Geometry} Single TES microcalorimeter design used in NETS. Here, the yellow square feature is the Au absorber, with the two shades of colors used to indicate the two separate Au layers. The TES bilayer consists of Mo (blue) and Cu (pink). The Cu banks, bars, and connection to the absorber in this same area are marked in orange. The entire design is sitting on a SiN absorber, with a frame of perforations encircling the absorber and sensor. Superconducting Mo traces run from both ends of the sensor to bond pads off the membrane at the edges of the chip, providing a route for the sensors to be connected to chips containing the bias circuit and SQUID amplifiers.}
\end{figure}

NETS contains an array of nominally identical TES x-ray microcalorimeters, the design of which is shown in Fig.~\ref{fig:TES_Geometry}. The design is an evolution of previous NIST designs, which use a sputtered MoCu bilayer for the TES material and evaporated Bi for the x-ray absorber\cite{hilton_microfabricated_2001, ullom_optimized_2005, swetz_current_2012, doriese_developments_2016}. Although these types detectors generally show roughly part in a thousand energy resolution, the evaporated Bi absorbers cause the detectors to suffer from non-Gaussian response functions ($\sim$20--30\% of x-ray events in low energy tails)\cite{yan_eliminating_2017}. It was found that Au absorbers could be used in place of the Bi absorbers to significantly reduce the low energy tail component of the detector response\cite{yan_eliminating_2017}. In the past designs, the Bi absorbers were evaporated directly on top of the TES thermistors. This cannot be done with the Au absorbers as Au is a metal (Bi is a semimetal) and would cause the TES superconducting properties to be intrinsically coupled to those of the Au absorber through the proximity effect\cite{werthamer_theory_1963, martinis_calculation_2000}, which is inconvenient from a design tuning and fabrication standpoint. Instead, the Au absorber is deposited laterally away from the TES thermistor in a ``sidecar'' design.

The sensor material is a sputtered MoCu bilayer with 215~nm of Cu on top of 65~nm of Mo. The thicknesses of the Cu and the Mo are tuned to target a specific superconducting transition temperature, $T_C$, via the proximity effect\cite{werthamer_theory_1963, martinis_calculation_2000}. The sample wafer is rotated during sputtering to maximize uniformity, and the sensors in this particular array have $T_C$ values in the range of 111--112~mK. This $T_C$ was chosen as a trade-off between the the theoretical energy resolution we could hope to achieve and cryostat hold time (see Sec.~\ref{subsec:Cryostat}). An additional Cu layer of thickness 419~nm is deposited on top of the sensor and patterned into banks, used to steer current toward the leads, and interdigitated stripes, used to control the transition shape and reduce excess noise\cite{lindeman_characterization_2004}. This Cu layer also thermally connects the sensor to the absorber. In addition to its use in the sensor bilayer, the Mo is used for superconducting leads and traces running off the membrane.

The absorber is made from Au that is electron beam evaporated in two separate layers, first 186~nm and then 779~nm thick, resulting in a total absorber thickness of 965~nm. The wafer is also rotated during this deposition and the evaporation tool had previously been characterized for roughly 3\% variation across a 76~mm wafer (better than 1\% variation across a single detector array chip). The absolute thickness had been measured with a stylus profilometer with $\lesssim$3\% uncertainty. Sec.~\ref{subsec:QE} contains a discussion of these absorber thickness uncertainties as they relate to the quantum efficiency. The deposition is done in two separate layers to reduce step coverage issues when coupling the absorber to the sensor through the Cu and to better control the thermal conductivity. The absorber area is 340~$\mu$m$\times$340~$\mu$m. 

When choosing the absorber thickness and area, trade-offs between dynamic range, quantum efficiency, and active area had to be made. In particular, the total absorber volume (sets the heat capacity and dynamic range) was chosen such that the detectors could operate in the linear region of their superconducting-to-normal transitions when measuring photons in the $\lesssim 15$~keV range. Beyond this region, the x-ray pulses would begin saturating and energy resolution would quickly degrade. The highest charge state that the NIST EBIT can produce is H-like Kr, with the brightest line at $\sim$13.5~keV, so we wanted to ensure that the detectors would be capable of measuring photons at these energies without saturating. The speed of the room temperature readout electronics (see Sec.~\ref{subsec:Readout}) imposes a limit on the slew rate of the pulses, which for these detectors occurs at roughly 10~keV. Reading out pulses above this $\sim$10~keV limit requires reducing the number of detectors to increase the readout speed of the remaining detectors and raise the slew rate limit. We note that the majority of our planned measurements are of highly charged ions with line energies below 10~keV. Once the saturation energy (total absorber volume) is chosen, we maximize the absorber area given the constraints of the array layout.

\begin{figure}
\begin{center}
\begin{subfigure}{\linewidth}
    \includegraphics[width=1.0\linewidth]{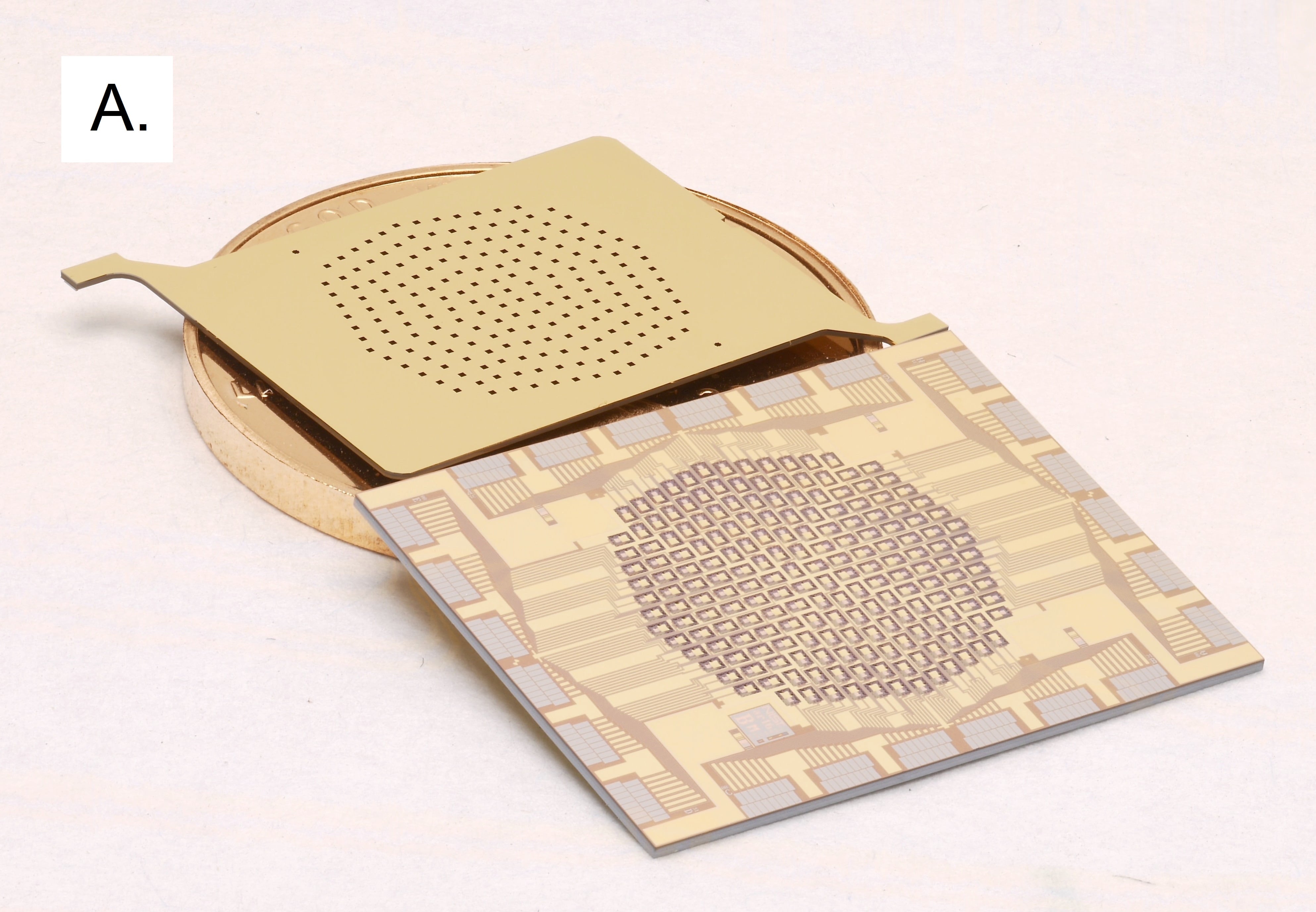}
\end{subfigure}
\begin{subfigure}{\linewidth}
    \includegraphics[width=0.975\linewidth]{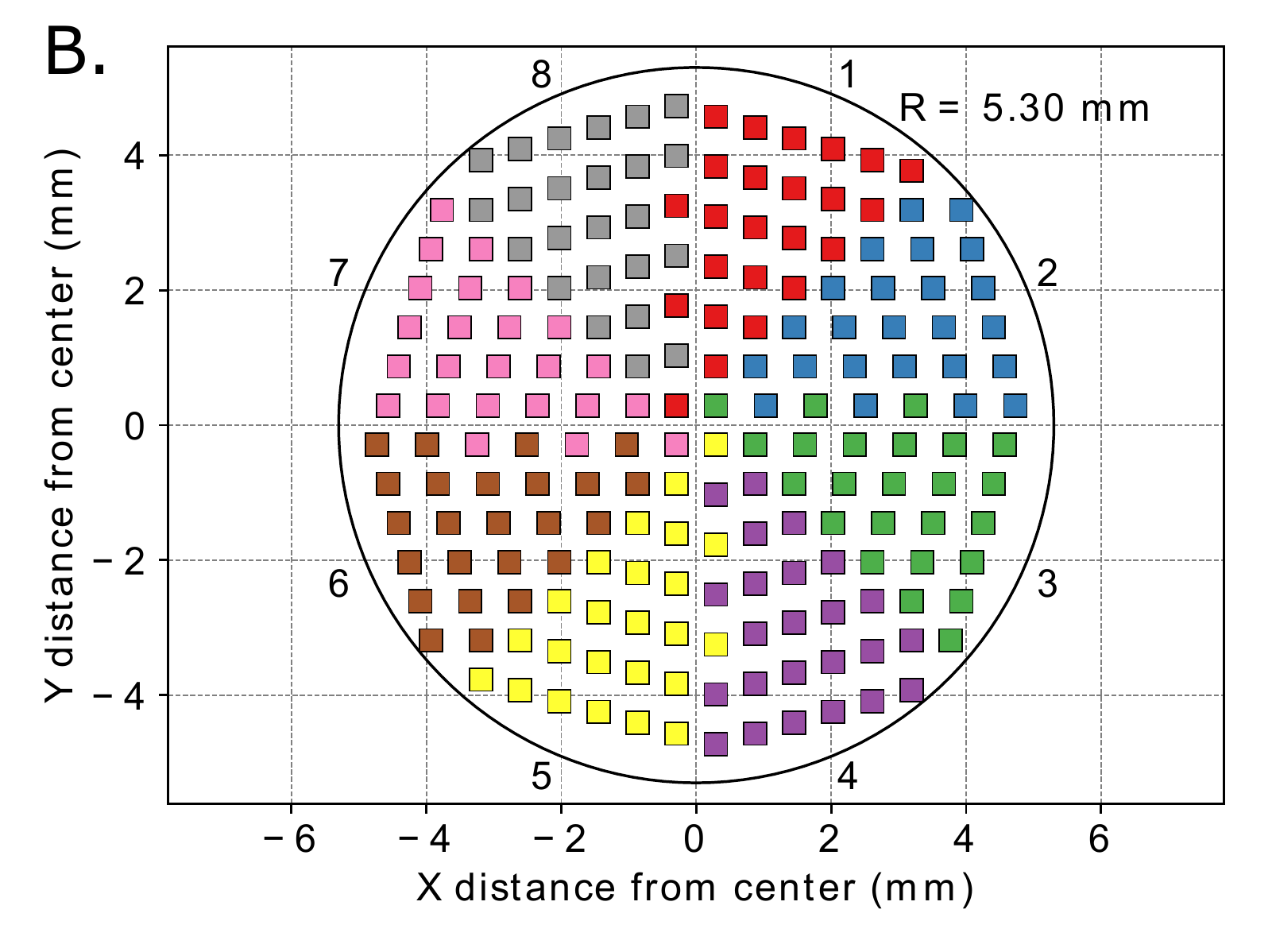}
\end{subfigure}
\end{center}
\caption{\label{fig:PixelMap} (A)~The TES array used in NETS. The aperture chip used to prevent x-rays from being absorbed by areas outside the TES absorbers is also shown (displayed on top of a penny, for scale). (B)~Pixel map indicating absorber size (rather than aperture size) and locations. The different colors represent the 8 readout columns, with 24 detectors in each column. A circle of radius 5.30~mm fully encloses all detector absorbers.}
\end{figure}

NETS houses a total of 192 TES pixels arranged in the roughly circular pattern depicted in Fig.~\ref{fig:PixelMap}. Here, 192 pixels are used instead of the 240 pixels that are used in the NIST evaporated Bi absorber design due to the increased pitch that is required when separated the absorber from the TES. Reducing the pixel count to 192 was required in order to keep the detector chip size constant and easily integrate it with our standard assembly package and readout chips. Out of fabrication and assembly, 166 of the 192 total possible detectors (86\%) are working and capable of detecting x-rays. Additional detectors may be flagged as bad in the data reduction steps depending on the data quality and restrictions of a particular analysis, as will be discussed in Sec.~\ref{sec:Reduction}.

\subsection{Readout}
\label{subsec:Readout}

NETS utilizes the TDM readout scheme with a two stage SQUID architecture\cite{doriese_developments_2016}. In this scheme, each detector gets its own first stage SQUID (SQ1) amplifier, and the signals from all detectors in a given readout column are coupled into a single second amplification stage SQUID array (SA). The SQ1s in a given column of detectors are addressed sequentially and have a maximum critical current of 10~$\mu$A. They are designed to have an asymmetric periodic response and are operated at the steeper of the two asymmetric slopes to maximize gain. The SAs contain 6 banks of 64 SQUIDs in a $2 \times 3$ (series~$\times$~parallel) configuration, with a maximum critical current of 60~$\mu$A. Here, the direction of the SA feedback coil current was reversed relative to TES spectrometers previously deployed by NIST, which was recently found to reduce SQUID amplifier instability and crosstalk between columns\cite{durkin_reducing_2019}. At room temperature, the signals are routed through a set of analog interface electronics and then through the digital feedback electronics\cite{reintsema_prototype_2003, doriese_time-division_2004}. The room temperature electronics orchestrate the TDM readout and maintain a separate flux locked loop for each SQ1 such that the feedback value is a measure of the TES current. In NETS, the detectors are divided into 8 parallel readout columns, with 24 rows (detectors) per column. NIST TDM multiplexers are described in greater detail in other publications\cite{reintsema_prototype_2003, doriese_time-division_2004, doriese_developments_2016}.

Data acquisition (DAQ) software is used to set SQUID and TES parameters, configure triggering conditions, and write raw pulse records (x-ray events). Typically, the SA bias current is set to $I_{c,max}$ (defined as the current which maximizes the amplitude versus input flux), while the SQ1 bias current is set to $\sim1.5\times I_{c,max}$ to reduce settling time after row switching. Other multiplexer parameters, such as those defining the regions of operation of the amplifiers, can also be tuned with the DAQ software. The TESs were biased in the superconducting transition at about 15\% of their normal-state resistance ($R_N$), which we found maximizes the resolving power of the detectors. More specifically, each readout column has a single bias line that is shared by all detectors within that column. The bias voltage for a column is set such that the column median detector resistance is at 15\%~$R_N$. The average variability in resistance at the bias point for a given column is 1.4\%~$R_N$. Generally, we see little if any degradation in resolution in detectors with bias points that differ from the column median by a few~\%~$R_N$  The DAQ software can also be used to decrease the number of columns and rows to read out, potentially lowering electrical crosstalk and improving the energy resolution at the cost of x-ray collection efficiency. Lowering the number of rows is also necessary when measuring high energy pulses ($\gtrsim 10$~keV), ensuring that the feedback loop of the room temperature electronics can keep up with the higher slew rates of these pulses.

The DAQ software is also used to configure an edge trigger for the collection of pulse data. Here, the edge triggering threshold (minimum slope in the feedback signal) is set far below the low energy cutoff in our quantum efficiency curve ($\sim$300~eV, see Fig.~\ref{fig:Efficiency}). With proper optimization, this triggering threshold can be set as low as a few 10s of eV (depending on desired $\sigma$ level of detector resolution), though this is rarely useful given the low energy cutoff in quantum efficiency. When taking noise measurements, the data is continuously streamed rather than run on a trigger. Raw pulse and noise records are written to disk in a binary format and are later analyzed using the data reduction pipeline described in Sec.~\ref{sec:Reduction}.

\subsection{Cryostat}

\label{subsec:Cryostat}

NETS utilizes an Adiabatic Demagnetization Refrigerator (ADR\cite{hagmann_adiabatic_1995, pobell_matter_2007}) to cool the detectors and SQUID electronics. This ADR was primarily chosen for its compact and portable design, and it is generally fairly simple to mate to the x-ray source\cite{doriese_practical_2017}. The cryostat consists of a large rectangular section containing the SA amplifiers and majority of the cooling hardware and a cylindrical protrusion, hereafter referred to as the snout, housing the detector package and SQ1 amplifiers. The bulk rectangular section is largely a commercially available component whereas the snout is custom built. Within the outer vacuum jacket are four thermal stages, each named after the nominal base temperature that stage reaches; these are the 50~K, 3~K, 1~K, and 50~mK stages. A pulse tube cryocooler is used to cool the 50~K stage to 50~K and the remaining stages to 3~K, eliminating the need for any liquid cryogens. Cooling the 1~K and 50~mK stages below 3~K is not continuous and involves a magnet ramp/deramp cycle (hereafter, ADR cycle). For this purpose, the 1~K stage contains a gadolinium gallium garnet (GGG) paramagnetic salt pill and the 50~mK stage contains a ferric ammonium alum (FAA) paramagnetic salt pill. A high field superconducting magnet is ramped to 3~T to align the magnetic moments in both salt pills while the stages are heat sunk to the 3~K stage. They are then thermally isolated from the 3~K stage, and the magnetic field at the salt pills is ramped down, a process which adiabatically disaligns the magnetic moments and requires them to absorb energy (heat from the salt pills). At the end of this cooling cycle, the NETS 1~K (GGG) stage reaches a base temperature $\sim$550~mK and 50~mK (FAA) stage reaches a base temperature of $\sim$45~mK. The ADR cycle is accomplished in roughly 1.5~hours.

In order to raise the temperature of the 50~mK stage from the base temperature to the detector operating temperature (here, 70~mK), we use the superconducting magnet to apply a small magnetic field ($\sim30$~mT). This field is controlled by a feedback loop to stabilize the temperature read out by the 50~mK thermometry to the operating temperature. The temperature of the 50~mK stage can be controlled at 70~mK for over 24~hours (hold time) before the various power loads on the stage cause the zero magnetic field temperature to exceed the operating temperature and the ADR cycle must be done again. Here, the operating temperature of 70~mK was chosen as a trade-off between detector performance and hold time. Generally we find improvements to detector performance through lower operating temperature to saturate when the operating temperature is $\sim$30~mK below $T_C$. Also, a hold time of around 24~hours is ideal for the typical work schedule at the NIST EBIT, allowing us to cycle the magnet at the start of the day and collect uninterrupted data with the spectrometer until EBIT operations end for the day. When controlling the 50~mK stage to 70~mK, we see a 4~$\mu$K standard deviation in the measured temperature. This is typical for systems we have deployed in the field\cite{doriese_practical_2017}, and we have found this level of stability to not limit single pixel energy resolution.

The array package (see Fig.~\ref{fig:TES_Operation}) is placed at the front of the 50~mK stage snout, simplifying the mating and minimizing the distance between the spectrometer and the x-ray source. Interface chips and SQ1 amplifiers are also placed in the 50~mK snout.  The 1~K stage cools the SA amplifier and associated circuitry.

Filters are positioned at the front of the 50~mK, 3~K, and 50~K snouts, which allow for transmission of x-rays while reflecting infrared (IR) radiation. These filters are clamped and glued into Al frames, which are then screwed into depressions at the front of the snouts and lined with Al tape at the edges to prevent IR light leaks. The main component of these filters is a thin ($\sim$110~nm), circular Al film with a diameter of 17.1~mm. The Al films on the 50~mK and 3~K snouts are free-standing, whereas the Al film on the 50~K stage is backed by a rectangular mesh composed of Ni. This mesh has a pitch of 363~$\mu$m with bars that are 15~$\mu$m thick and 30~$\mu$m wide. Al was chosen as the film material due to ease of fabrication and because it is a low-Z material with a high x-ray transmittance in the NETS energy range.  

We expect the filter on the 50~K stage to reflect $\gtrsim$99\% of the 300~K radiation on this stage\cite{ordal_optical_1988}, and we expect higher reflectance of the 3~K and 50~mK filters to 50~K and 3~K radiation, respectively. The remainder of the IR radiation is expected to be absorbed by the filters, which can heat the filters and lead to increased blackbody radiation. For this reason, the filters need to be thick enough to thermalize well with the surrounding stage and not cause excess thermal loading to the innermost stage. Excess thermal loading can degrade the energy resolution of the TES pixels. The 50~K stage filter, which is exposed to more radiation than the other filters, includes a Ni mesh to improve thermalization. A thicker ($\sim$200~nm) Al film without a Ni mesh could be used in the case that the Ni mesh limits our ability to reach a target uncertainty in a line intensity measurement. One reason we attempted to maximize low energy transmission is the possibility to take simultaneous measurements on lines with energies in this region with NETS and with an existing extreme ultraviolet (EUV) spectrometer\cite{gillaspy_visible_2004} deployed at the NIST EBIT, which is limited to energies of up to $\sim$500~eV. 

A similar x-ray window is placed at the front of the outermost snout (vacuum snout), but this one is also capable of holding atmospheric pressure and has a larger diameter (25.4~mm). Here, the Al film is $\sim$65~nm thick and is backed by a $\sim$700~nm thick polyimide film with a by weight composition of 3\%~H, 71\%~C, 8\%~N, and 18\%~O. This window also contains a roughly hexagonal mesh composed of 304 stainless steel. This mesh is on a 360~$\mu$m pitch and has bars that are 100~$\mu$m thick and 30~$\mu$m wide. The errors in the filter/window parameters as they relate to the system quantum efficiency are discussed in Sec.~\ref{subsec:QE}.

\section{Integration with the EBIT}
\label{sec:Integration}

\subsection{Overview}

\begin{figure}
\begin{center}
\begin{subfigure}{\linewidth}
    \includegraphics[width=0.9\linewidth]{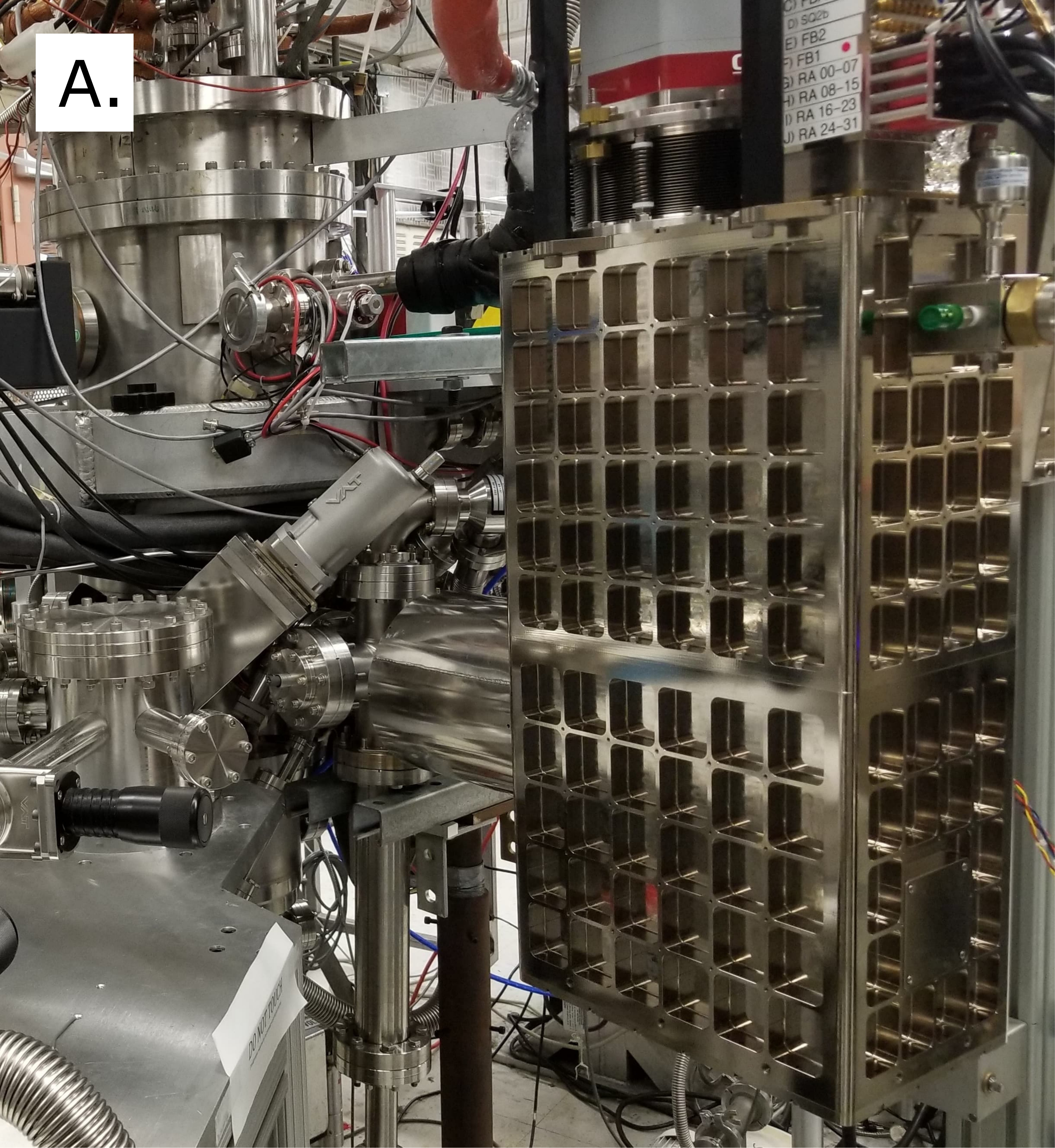}
\end{subfigure}
\par\medskip
\begin{subfigure}{\linewidth}
    \includegraphics[width=0.9\linewidth]{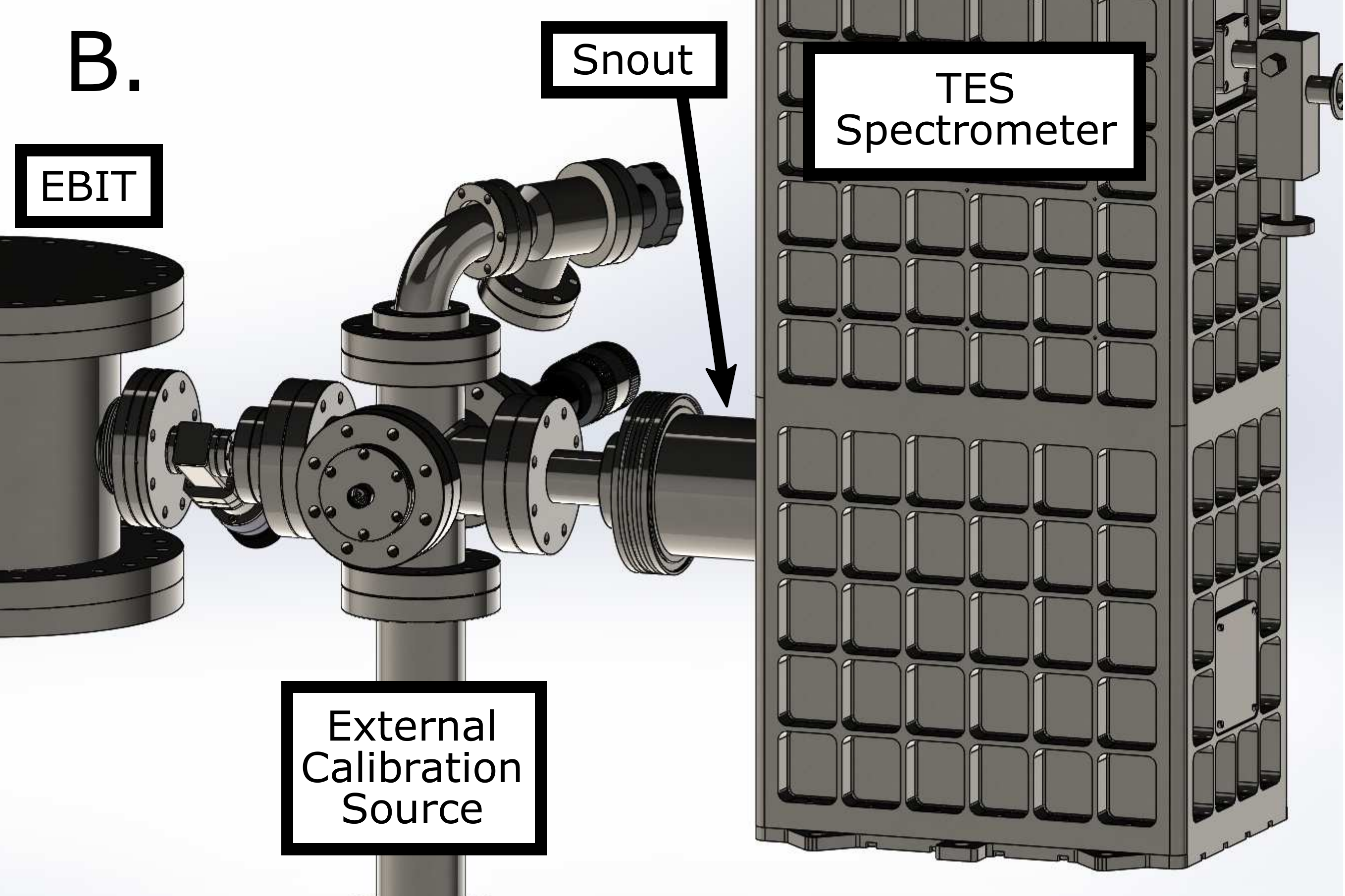}
\end{subfigure}
\end{center}
\caption{\label{fig:SystemIntegration} (A)~Photograph and (B)~Computer aided design (CAD) rendering of the TES x-ray spectrometer mounted to the NIST EBIT. The spectrometer (right) is mated to a viewport in the midplane of the EBIT (left) through a 6-way vacuum cross that also provides access to an external x-ray calibration source (center). In the CAD rendering, the seven remaining EBIT viewports mated to other detectors and EBIT components as well as external magnetic shielding around the spectrometer snout are not pictured, for clarity.}
\end{figure}

NETS was integrated with the EBIT as shown in Fig.~\ref{fig:SystemIntegration}. The spectrometer is held up with a rigid stand through a single point hanging connection at the top of the cryostat and a set of screws to fix the rotation. The stand is used align the spectrometer to the viewport of the EBIT and hold up a fraction of the weight of the external calibration source after the spectrometer has been fixed to the EBIT. A remote motor (not pictured) for the pulse tube cryocooler was placed on a rigid stand at the same height as the pulse tube cryocooler, with a He gas hose connecting the two components at roughly a 90$^{\circ}$ angle to minimize vibrations at the cryostat. 

\begin{figure*}
\begin{center}
    \includegraphics[width=1.0\linewidth]{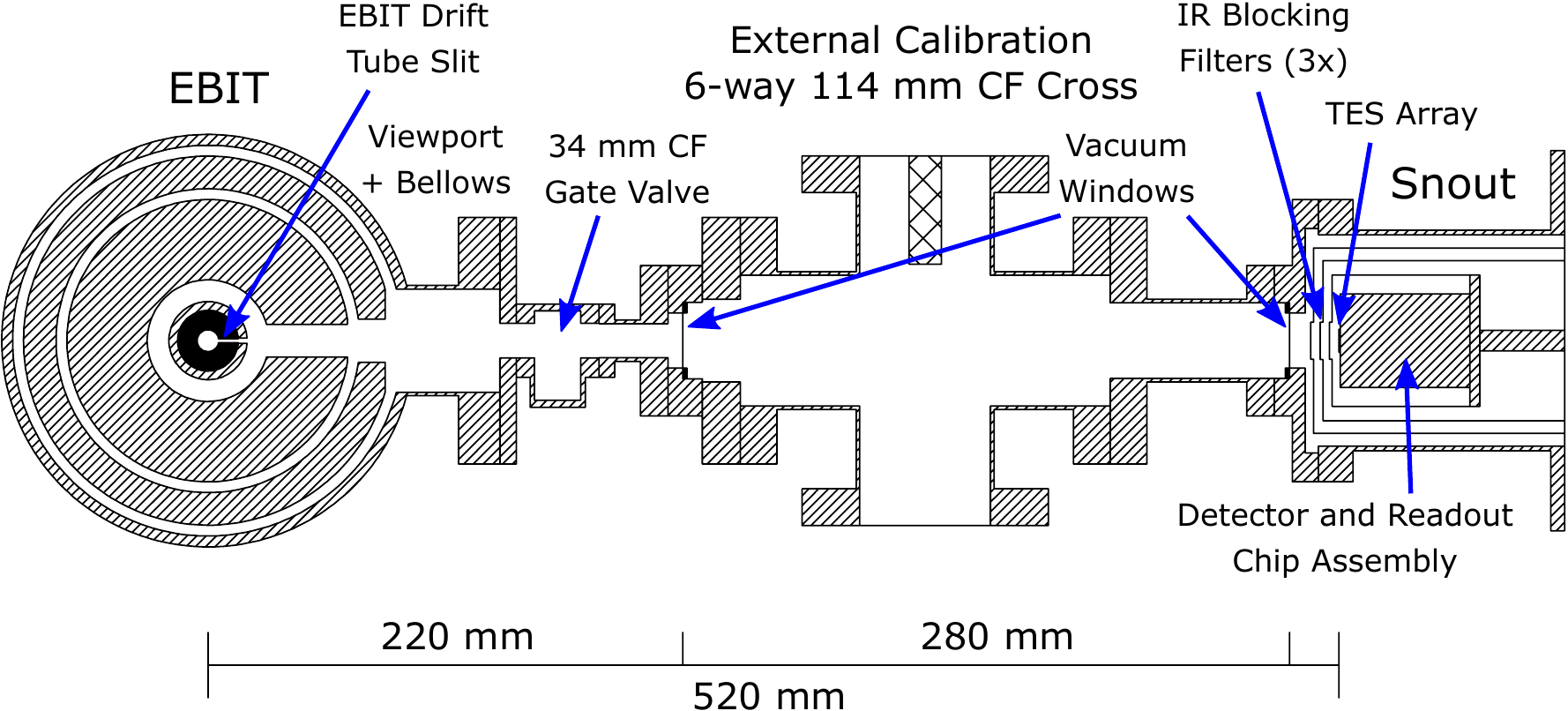}
\end{center}
\caption{\label{fig:OpticalPath2D} Top-down cross-section schematic of NETS mounted to the NIST EBIT through an external calibration source chamber. The total distance between the EBIT trap center and the TES array is 520~mm, with vacuum windows positioned 220~mm and 500~mm away from the EBIT trap center. IR blocking filters internal to the spectrometer cryostat are positioned $\sim$4~mm apart, with the innermost 50~mK filter positioned $\sim$4~mm from the TES array. Note that the solid angle is limited by the radius of the active area of the array and is not obstructed by other components along the x-ray beam path, such as the drift tube slits or gate valve walls.}
\end{figure*}

The front of the TES snout is mounted to a 6-way cross that houses retractable metal targets for an external calibration source (Sec.~\ref{subsec:CalibrationSource}). At the opposite end of this 6-way cross mates to a viewport in the midplane of the EBIT through a second vacuum window (same designed composition as window at front of TES snout), mechanical valve, and short bellows (16~mm length). A top-down cross-section of this arrangement is shown in Fig.~\ref{fig:OpticalPath2D}. The current setup therefore includes three separate vacuum spaces during regular operation: the NETS cryostat, the intermediate area between the cryostat and the EBIT that includes the calibration source, and the EBIT chamber. The separate vacuum space of the NETS cryostat was used to ensure a lower base pressure in the EBIT chamber. Here, the cryostat is designed only for high vacuum (HV) and uses Klein Flange (KF) fittings whereas the calibration source and EBIT chamber are designed for UHV with ConFlat (CF) fittings. The second vacuum window between the EBIT chamber and calibration source was used as an additional precaution during the original commissioning run. In the case of a vacuum failure in the calibration source chamber during operation of the EBIT and NETS, the conditions in the EBIT chamber and spectrometer cryostat would be preserved. After gaining some confidence with our setup, one of these vacuum windows may be removed to improve collection efficiency (still need one vacuum window to separate HV cryostat side from UHV EBIT side). With the current configuration, the typical operating pressure in the EBIT is $\lesssim 10^{-8}$~Pa ($10^{-10}$~Torr) when the field coils are running and actively cooled with liquid He.

\subsection{X-ray Collection Efficiency}
\label{subsec:QE}

\begin{figure}
\includegraphics[width=1.0\linewidth]{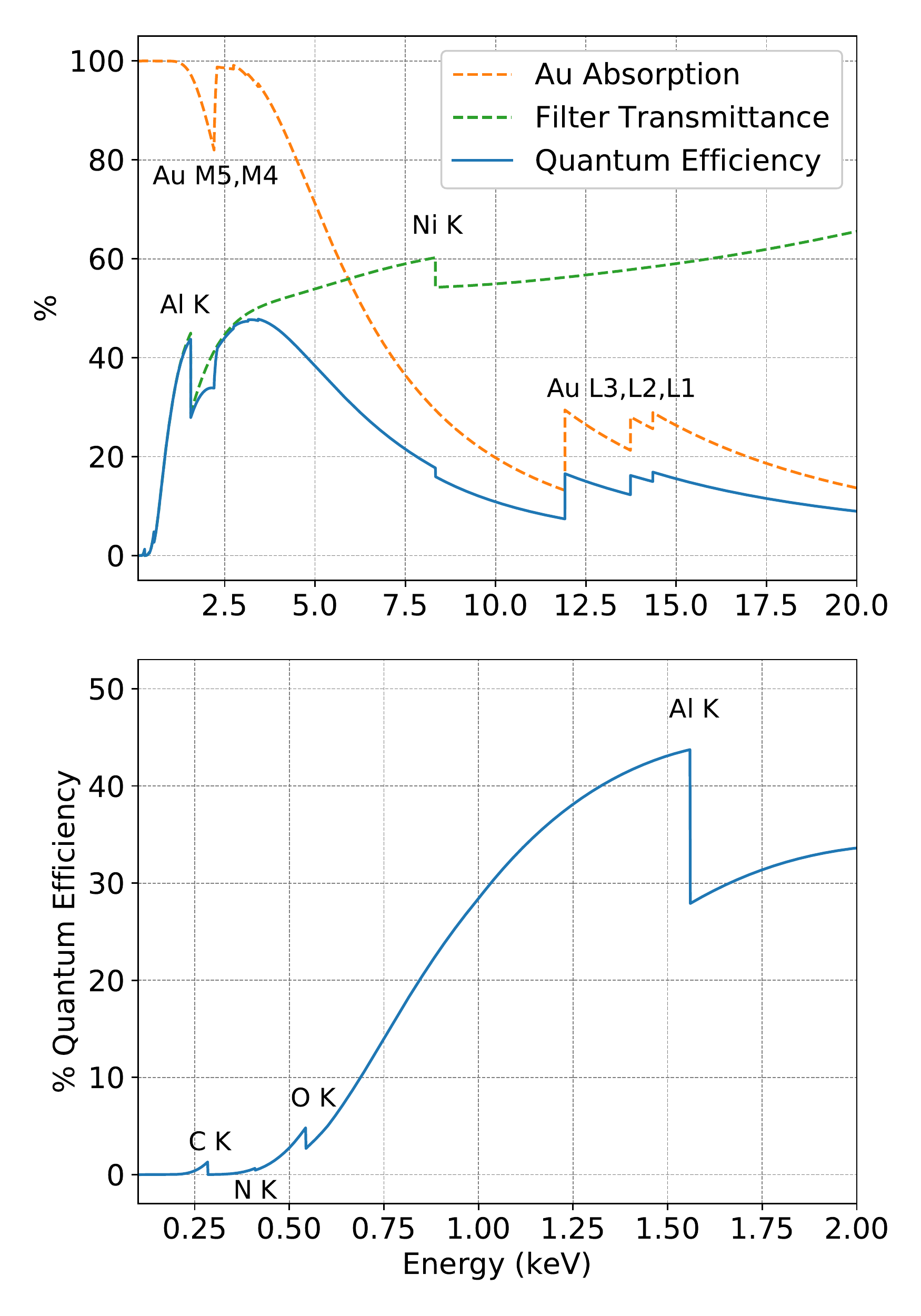}
\caption{\label{fig:Efficiency} \textit{Upper}: Calculated average QE for a detector in NETS, including losses due to transmittance through the 2 vacuum windows and the 3 IR-blocking filters as well as the properties of the Au absorbers. The absorption curve is labeled with the M and L absorption edges of the Au absorbers and the filter transmittance curve is labeled with the the most prominent absorption K edges of materials in the filters, primarily Al (from film) and Ni (from mesh). The reduction in collection efficiency due to the aperture chip masking of the absorber area ($\sim$32\%) is not included in this figure. \textit{Lower}: Low energy portion of the QE curve, which is dominated largely by the filter transmittance. In addition to the Al K edge, the K edges of C, N, and O (from polymer backing) are visible here.}
\end{figure}

An understanding of the quantum efficiency (QE) curve and associated uncertainties will be especially important when the scientific topic of interest is the measurement of line intensity ratios. Using the xraylib\cite{schoonjans_xraylib_2011} library and thickness and elemental density measurements of the windows and filters, we calculated their expected transmittance. We combined this transmittance with the expected absorption of a 965~nm thick Au absorber to calculate a total detector QE. The average filter transmittance, Au absorption, and combined QE curves are shown in Fig.~\ref{fig:Efficiency}. Here, different detectors will have varying QE curves primarily due to the geometry of the mesh structures supporting the 50~K filter and vacuum windows. In calculating the transmittance curve in Fig.~\ref{fig:Efficiency}, the transmittances were averaged across the filter. The QE has its peak of 47.8\% at an x-ray energy of 3.43~keV. 

As for the uncertainties in these parameters, the Au absorbers have an absolute thickness uncertainty of $\lesssim$3\% with $\lesssim$1\% thickness variation across the array (see Sec.~\ref{subsec:DetectorArray}). In the filters and windows, the uncertainty in the absolute Al thickness is 2\% in the IR-blocking filters and 3\% in the vacuum windows. The uncertainty in the elemental area densities of the polyimide is 2\%. Both the Al film and polyimide display minimal spatial variation. The combined uncertainty from these components is fairly small ($<$1\% $\Delta$QE at 3.43~keV, including pixel-to-pixel variations), but the uncertainty added by the stainless steel meshes on the vacuum windows and the Ni mesh on the 50~K filter is more significant.

The stainless steel meshes on the vacuum windows are fairly thick ($\sim$100~$\mu$m), and x-ray transmission through the mesh is expected to be low ($<$1\% up to $\sim$15~keV). The main uncertainty here comes in the open area of the mesh (80--82\%), and although this will create pixel-dependent variations in the illumination, due to the thickness of mesh, these variations are not expected to be energy-dependent. The Ni mesh on the 50~K filter similarly has an uncertainty associated with the open area (82--84\%) of the mesh. In addition, the Ni mesh is relatively thin (15$\pm$2~$\mu$m), and there can be fairly large transmission of x-rays through the mesh, especially near the K absorption edge (up to $\sim$50\%) and above $\sim$10~keV ($\sim$5--65\% between 10--20~keV). This will cause energy-dependent variations in illumination across different pixels on the array. When measuring line intensity ratios, these effects are in part mitigated by coadding (averaging) results across all detectors. If necessary for a given application, a series of measurements at various x-ray energies can be taken to map out the mesh positions relative to the TES microcalorimeters, given an expected no-mesh illumination pattern.


The distance between the face of the TES array and the trap center is nominally 520~mm when the system is under vacuum, with 1--2~cm of leeway due to contraction of the bellows and exact positioning of the array package along the 50~mK stage during assembly. The active area of the array fits within a 5.3~mm radius circle, but only x-rays that hit an absorber are accurately detected by the spectrometer. A collimator with an array of 280~$\mu$m$\times$280~$\mu$m apertures, slightly smaller than the absorber length and width, is placed over the detector array to ensure x-rays making it to the array only hit the absorbers. The area of the individual apertures is smaller than the detector absorber area (340~$\mu$m$\times$340~$\mu$m) in order to account for any inaccuracy in alignment when mounting the aperture chip. The active area of the array is therefore reduced to 15.1~mm$^2$, down from the 22.2~mm$^2$ that would be expected from the total area of the absorbers ($\sim$32\% reduction). This results in a solid angle of $5.57 \times 10^{-5}$~sr = 0.183~deg$^2$. In other words, assuming isotropic emission, a fraction $4.43 \times 10^{-6}$ of x-rays originating from the trap center can make their way to a detector, and the chance that one of these x-rays is actually absorbed by a detector is determined by the QE curve in Fig.~\ref{fig:Efficiency}. It should be noted that the idea of solid angle may not be perfectly correct here as the x-ray emitting region is more an extended line source rather than a simple point source. The beam diameter is $<100$~$\mu$m, and a roughly 20~mm length of the trap is visible through the vertical drift tube slits. Based on this beam shape, the distance of the source to the detectors, and radius of the active area of the array, we would expect a uniform x-ray illumination of the pixels on the array if the meshes backing some of the filters and windows were not present.

\subsection{Calibration with External X-ray Sources}
\label{subsec:CalibrationSource}

We plan to use well known reference lines to generate calibration curves for the NETS microcalorimeters. Although the EBIT is capable of generating many such lines, it is limited to the gas and metal targets accessible to the EBIT at the time of measurement, and exchanging these targets can be a lengthy process. This leads to regions in energy space in which the EBIT cannot generate lines with well known positions, which can be problematic if a measurement requires energy calibration in such a region. A more flexible method for generating well known x-rays lines for calibration was desired.

For this reason, there is a 6-way cross between the EBIT and spectrometer used to mate an external x-ray calibration source to the system. This source is composed a of set of solid samples (targets) and a mechanism to excite characteristic radiation from those targets. We can directly excite conductive targets through electron impact excitation with an electron gun, or create secondary x-rays with conductive or non-conductive targets through illumination with a commercial x-ray tube source. We have used two tube sources, with Pd and W anodes, and a variable beam energy of up to $\sim$25~kV and current of up to $\sim$2~mA. The targets are on a linear actuator, allowing for manual switching of the targets without needing to break vacuum. The targets are held at 45 degrees to the electron gun and to the spectrometer, enabling characteristic x-rays to efficiently arrive at the spectrometer. This flexibility in source type and targets is useful for generating characteristic x-rays and calibration curves targeted for a specific application.

\begin{figure}
\begin{center}
\begin{subfigure}{\linewidth}
    \includegraphics[width=1.0\linewidth]{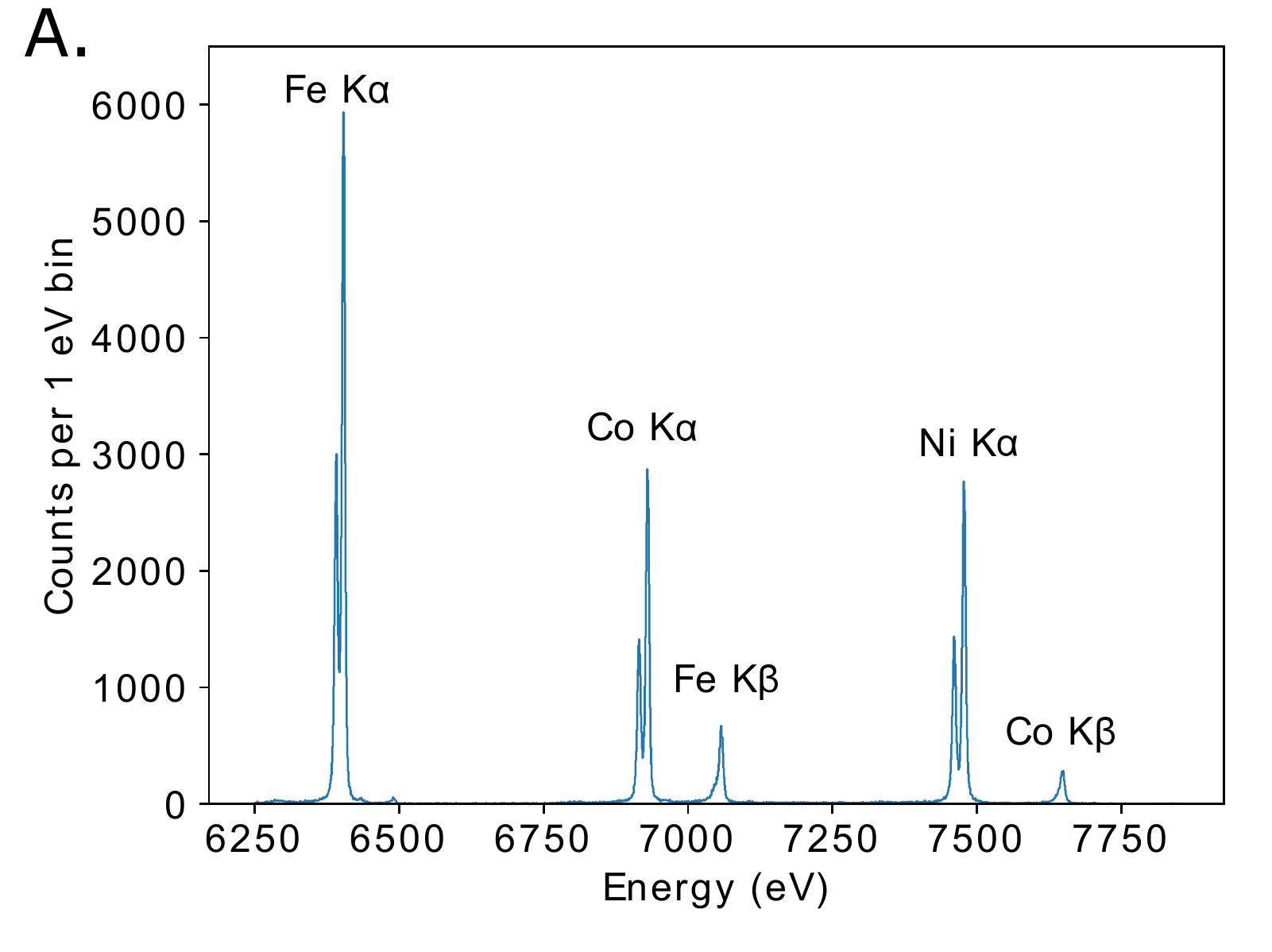}
\end{subfigure}
\par\medskip
\begin{subfigure}{\linewidth}
    \includegraphics[width=1.0\linewidth]{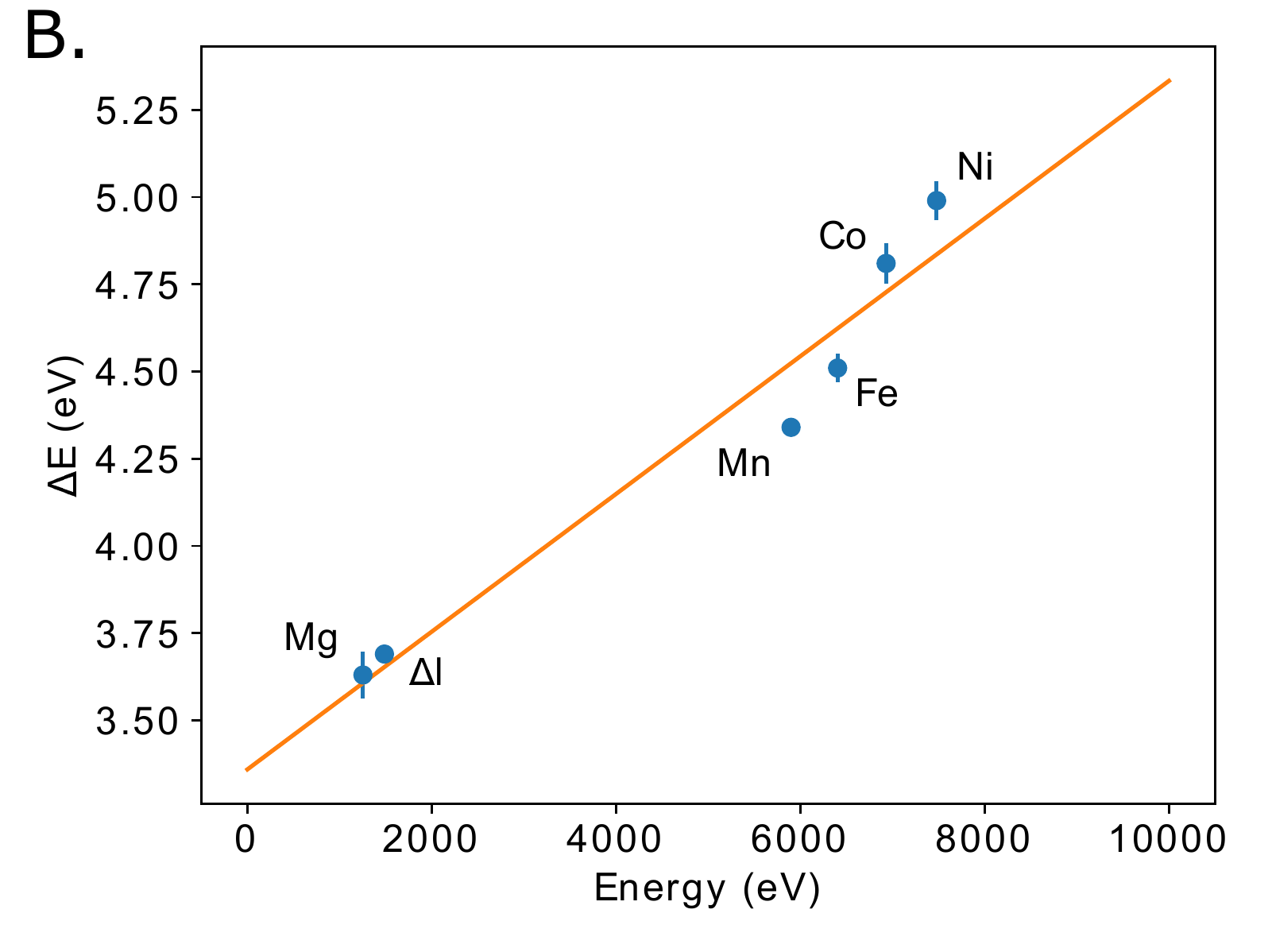}
\end{subfigure}
\end{center}
\caption{\label{fig:CalibrationSource} (A)~Spectrum from example external calibration source measurement. Here, Fe, Co, and Ni were used as the targets and the total array count rate was kept to under 200~cps. The K$\alpha$, and to a lesser extent K$\beta$, lines are the strongest features observed in the spectrum. (B)~Detector energy resolutions extracted from fits to K$\alpha$ lines from a variety of external calibration source measurements. The resolution is measured to be 3.7~eV at the Mg~K$\alpha$ (1.25~keV) and 5.0~eV at Ni~K$\alpha$ (7.48~keV). The best fit line is used to show the expected energy resolution of the detectors across this energy region. Here, only the statistical uncertainty associated with fitting the line model is shown. Other sources of uncertainty, including uncertainty in the line shape model, slight variations in noise pickup and electrical crosstalk, and excess drift leftover after correction (see Sec.~\ref{sec:Reduction}), account for a scatter of $\sim$100~meV.}
\end{figure}

A spectrum taken using the external calibration source with a target consisting of Fe, Co, and Ni foils is shown in the top panel of Fig.~\ref{fig:CalibrationSource}. X-rays were excited with an x-ray tube source with W anode. A beam voltage of 15~kV was used, and the current ($\sim$20~$\mu$A) was tuned to keep the count rates at $\sim$1~cps/detector in order to resemble the low rates expected in the EBIT measurements. The data analysis is described in Sec.~\ref{sec:Reduction}. The energy resolution extracted from K$\alpha$ lines in the coadded spectra of this and other external calibration source measurements is shown in the bottom panel of Fig.~\ref{fig:CalibrationSource}. Here, the expected natural line shapes\cite{klauber_magnesium_1993, schweppe_accurate_1994, holzer_pre$kensuremathalpha_12$/pre_1997} are convolved with a Gaussian representing the detector response, and the Gaussian FWHM is taken to be the energy resolution.

\subsection{External Magnetic Shielding}

The low temperature TES array and SQUID electronics are sensitive to magnetic fields. Excess magnetic fields could raise the $T_C$ of the superconducting components, broaden the superconducting-to-normal transition in the detectors, and cause shifts in SQUID amplifier gain. In worst cases, magnetic flux trapped in the superconducting components can drive them into their normal state, rendering detectors or SQUIDs with trapped flux unusable until the flux traps are released by warming the components far above their $T_C$, a lengthy process. For this reason, in previous NIST TES spectrometers the 50~K and 3~K snout shields enclosing these components are comprised of mu-metal\cite{jiles_introduction_1998} to attenuate Earth's magnetic field, typical background laboratory fields, and fringing fields from the ADR magnet (snout magnetic field shielding described further in Ref.~\citenum{doriese_practical_2017}). In addition, the 50~mK snout and filter enclosing the TES array and first stage SQUIDs are fabricated with Al, which goes superconducting during an ADR magnet cycle and provides additional magnetic field shielding\cite{hamilton_superconducting_1970}. Similarly, the SA amplfiers on the 1~K stage are enclosed by a superconducting Nb box. These components are also shielded from the internal ADR magnet fringing fields through a cylindrical vanadium permendur magnet enclosure (for a detailed description of ADR magnet shielding with vanadium permendur, see Ref.~\citenum{bennett_high_2012}). Extrapolating from measurements on systems with similar shielding arrangements\cite{bennett_high_2012, doriese_practical_2017}, we can expect a magnetic field on order 1~$\mu$T in the area of the detector package when the ADR magnet is controlled at fields that are typical for temperature regulation. Static fields of this magnitude are largely accommodated for by careful gradiometric SQUID amplifier design\cite{stiehl_time-division_2011} and proper selection of the SQUID biasing parameters. Drifts in the amplifier gain due to this field decaying over the course of an ADR cycle have been measured to be small compared to thermal drifts and are corrected for during pulse processing, as will be discussed in Sec.~\ref{sec:Reduction}.

Somewhat unique to this setup was the additional strong stray magnetic field (primarily in the transverse direction of the snout) of the main EBIT magnet in the region of the NETS snout. The compactness of the various snout temperature stages does not allow for additional interior magnetic field shielding without a costly redesign of the snout. We instead found it much more simple to mount two concentric 1.6~mm thick cylindrical mu-metal shields (designed for factor of $\sim$100 attenuation in the transverse direction) over the vacuum snout where space is much less limited. The goal here was to reduce the expected magnitude of the stray EBIT field directly outside the vacuum snout to levels comparable to typical laboratory background. The shielding internal to the cryostat could then null the remaining field at the detectors and SQUIDs to values typically seen under temperature regulation. We tested the magnetic field suppression of these external shields using a 3-axis Hall effect magnetometer with an estimated uncertainty of $\lesssim$5\%. When the EBIT is ramped to a maximum field current of 147.7~A, we measure a field of 3~mT directly outside the external mu-metal shield and 30~$\mu$T between the external mu-metal shield and vacuum jacket near where the array is located. This is below the background field in the laboratory, which was measured to be 58~$\mu$T prior to ramping the EBIT magnet. With this, we expect no significant excess field in the area of the detectors and SQUIDs from the EBIT magnet.

\section{Measurements}
\label{sec:Measurements}

For our initial tests we took x-ray EBIT measurements on previously studied ions to demonstrate the capabilities of NETS. These are summarized in Table~\ref{tab:MeasurementLog}. The majority of these measurements were done on He-like and H-like ions of injected neutral gases, including CO$_2$, Ne, and Ar. We also used the MeVVA\cite{holland_low_2005} ion source of the NIST EBIT to inject metallic ions into the trap. Here, we primarily focused on Ni-like W, which was the subject of a recent EBIT study\cite{clementson_spectroscopy_2010}. The ions listed in Table~\ref{tab:MeasurementLog} have fairly well known line shapes and positions, useful for benchmarking a new instrument. They can also be used as calibration lines for measurements of ion charge states with less well known spectra, which is planned in future measurements.

\begin{table*}
\caption{\label{tab:MeasurementLog} Measurement log. The ions studied in this intial measurement campaign were He-like and H-like ions of injected neutral gases and Ni-like ions of metallic W injected with a MeVVA.}
\begin{ruledtabular}
\begin{tabular}{cccccc}
Ion & Shield & Collector & Ionization & Count Rate & Total \\
 & Voltage (kV) & Current (mA) & State & (cps/array) & Counts \\
\hline
Ne     & 4.0  & 99  & He/H-like & 180.9 & 325736 \\
Ar     & 10.0 & 92  & He/H-like & 51.4  & 185935 \\
CO$_2$ & 2.0  & 53  & He/H-like & 35.5  & 139690 \\
\hline
W      & 4.1  & 105 & Ni-like   & 136.4 & 651124 \\
W      & 4.1  & 105 & Ni-like   & 164.1 & 406991 \\
Ar     & 9.5  & 108 & He/H-like & 51.3  & 192346 \\
Ne     & 4.0  & 108 & He/H-like & 179.6 & 343093 \\
\hline
Ne     & 4.0  & 139 & He/H-like & 266.1 & 577873 \\
W      & 4.1  & 137 & Ni-like   & 264.1 & 943025 \\
Ar     & 9.5  & 140 & He/H-like & 52.4  & 127000 \\
W      & 4.1  & 145 & Ni-like   & 232.4 & 378798 \\
CO$_2$ & 2.0  & 58  & He/H-like & 58.8  & 79118  \\
\end{tabular}
\end{ruledtabular}
\end{table*}

We report on measurements taken across three days, and some details of the experiment varied day to day. The horizontal lines in Table~\ref{tab:MeasurementLog} separate measurements taken on different days. The NETS ADR magnet was cycled at the beginning of each morning, altering the magnetic field environment in the detectors and SQUIDs and changing gain values. For this reason, we take calibration data for the TES microcalorimeters each day, and data from different days of measurement are analyzed independently from one another. We note that precise ($\sim\pm$0.1~eV) measurements of line energy require more calibration data than other measurements. We have not investigated whether the calibration is stable enough to be reused from day to day in any case. The ion trap dumping/recycling time was set at 5~s for the first day of measurements, but then was subsequently changed to 3~s on day 2 and 3. This is a parameter that gets tuned using past experience operating the EBIT in order to improve charge state purity and increase count rates. The Fe stand originally supporting the external calibration source was found to have a permanent magnetic field which limited the maximum collector current (proxy for the maximum count rate at the spectrometer) of the EBIT to below past levels. This stand was exchanged for an Al stand between days 2 and 3, after which the collector current was identical to that achieved prior to external calibration source installation. The higher maximum collector current resulted in higher count rates on day 3, as shown in Table~\ref{tab:MeasurementLog}. With respect to detector performance, these are relatively low count rates ($\lesssim 2$~cps/detector), so pulse pileup\cite{fowler_microcalorimeter_2015} and crosstalk\cite{durkin_reducing_2019} events happen infrequently and have very little effect on the data quality.

\begin{figure}
\includegraphics[width=1.0\linewidth]{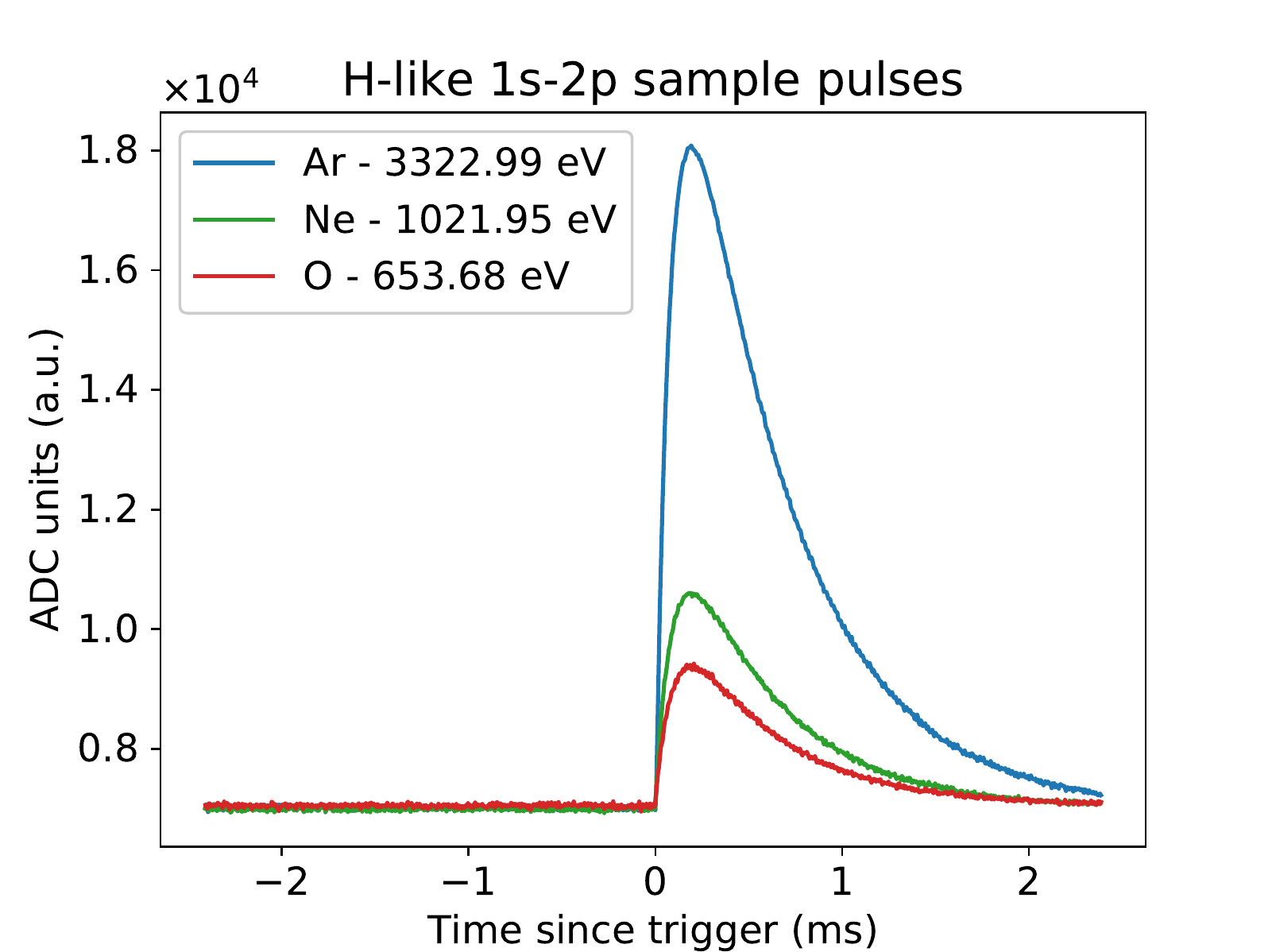}
\caption{\label{fig:Traces} Sample pulse records for a single microcalorimeter from the H-like 1s-2p transition in O, Ne, and Ar. Here, the transitions occur at three different energies, and the taller pulses indicate more energetic transitions. Note that to get a precise estimate of the x-ray energy, further pulse processing and filtering is required.}
\end{figure}

The raw data products are pulse records, each containing the information of a single x-ray event. Some sample pulse records from the H-like 1s-2p transition in O, Ne, and Ar are shown in Fig.~\ref{fig:Traces}. The height of a trace is representative of the energy of the incident x-ray, but for a more exact determination of the energy, further pulse processing must be done, including combining the pulse records with noise data and optimally filtering the pulses (steps outlined in Sec.~\ref{sec:Reduction}). The pulses across all working microcalorimeters have an average 1/e rise time of 60~$\mu$s and fall time of 640~$\mu$s. The pulse record length was chosen to be long enough to store all the information in a pulse needed to accurately determine the pulse energy, but not too long as to reduce throughput due to pulse pile-up and detector crosstalk. For these reasons, a record length of 1000 samples (4.8~ms) was chosen, with 50\% of the record allocated to the pre-trigger (samples before x-ray arrival).

\section{Data Reduction}
\label{sec:Reduction}

We used the Microcalorimeter Analysis Software System (MASS) Python package to reduce data from raw pulse and noise records to a variety of output products such as energy calibrated spectra\cite{fowler_practice_2016, becker_advances_2019}. First, the raw data pulses (e.g., Fig.~\ref{fig:Traces}) are passed through a set of cuts designed to remove piled up pulses and other irregularities. An average pulse profile is then created for each detector, which is used along with noise data to optimally filter the pulses. Next, the pulses undergo a temporal drift correction. Finally, the detectors are energy-calibrated using well known x-ray lines. The final product is a list of time-tagged, energy-calibrated x-ray events for each detector. The steps of this analysis routine are discussed in more detail below.

Directly after loading the raw pulse records, a variety of summary quantities are calculated. Two of these quantities, the pre-trigger root mean square (RMS) and post-peak derivative, are used to cut out anomalous pulses. An unexpectedly high pre-trigger RMS typically indicates that the detector was triggered during the decay of a prior x-ray pulse, whereas a high positive post-peak derivative likely indicates a second x-ray arrived after the peak of the triggering pulse. Additionally, we use outlier rise and peak times as indicators of secondary pulses arriving during these periods, though these events are fairly rare within these short periods. Noise record data are aggregated to determine expected deviations for these cut quantities, setting the limits for these cuts. Typical pulse cut fractions were less than 2\% at these low count rates. These are high SNR cut quantities, so we do not see any energy dependence to these cuts. The exception to this is when an x-ray is energetic enough to produce a pulse with a slew rate that exceeds the limit set by the speed of the readout electronics, resulting in a highly abnormal pulse record that is generally cut for a combination of the summary quantities listed above. In an energy spectrum, this would be seen as a drop off in counts near this high energy readout limit. In addition to cutting pulses from detectors, entire detectors may be cut during this step. Here, we check to see if any detectors have an abnormally low number of accepted pulse records. This typically indicates a more serious issue with the individual detector or the readout chain of that detector.

The next step in the data reduction pipeline is optimally filtering the pulses\cite{szymkowiak_signal_1993, anderson_optimal_2005}. First, an average pulse is created for each detector using records at a line(s) of interest in the raw pulse height spectrum, although we find the latter steps of the data reduction to be largely insensitive to the exact choice in energy at which this average pulse is created. The average pulse is combined with the noise data, which is assumed to be stationary, in order to produce a filter that maximizes the SNR when estimating the pulse height. This filter is designed to be insensitive to the exact arrival time of a pulse relative to the sampling clock\cite{fowler_practice_2016}. The noise data are typically taken during the same day as the measurements (same ADR cycle), largely based on precedent; we measure roughly the same energy resolution when using noise data from another day. Each pulse record is passed through the filter for its associated detector and a filtered pulse height is extracted.

In NETS, we observe a $\sim (3 - 5) \times 10^{-4}$/hr drift in the fractional pulse height, which varied from detector to detector. A major portion of this drift results from the fact that the 1~K stage temperature increases from $\sim$500~mK immediately after an ADR cycle to $\sim$1~K 10 to 20 hours later. To reach line placement accuracy to better than a few parts in $10^4$ for measurements with integration times of on order hours it is necessary correct for this drift. We use a spectral entropy-minimizing algorithm that uses the baseline, or pretrigger mean, and the x-ray timestamps as indicators of drift in the filtered pulse heights (see Ref.~\citenum{fowler_practice_2016} for more detail). Typically, one or multiple strong lines in the calibration data are measured at multiple times during an ADR cycle to generate the data necessary for drift correction, but strong lines in the measured EBIT data could also be used for this purpose if they exist for a given measurement. This approach to drift correction has previously been used to achieve on order $10^{-5}$ fractional line position accuracy\cite{tatsuno_absolute_2016, fowler_reassessment_2017}. In addition to the drift in pulse height throughout an ADR cycle, there is also a small amount of pulse height variation between different ADR cycles (detectors typically show $\sim 10^{-4} - 10^{-3}$ cycle-to-cycle variation in fractional pulse height when measured at the same time after the start of an ADR cycle). We generally calibrate and analyze data from different ADR cycles more or less independently, so these cycle-to-cycle changes are far less significant to the analyzed data quality than the drift observed during an individual ADR cycle.

Energy calibration is performed by identifying a set of well known lines in a particular spectrum, fitting those lines with their known lineshape and the detector response function, and generating an interpolating function between filtered pulse height (including the various corrections) and energy. We identify lines in the filtered pulse height spectrum of a single arbitrarily chosen reference pixel by hand, then use a dynamic time warping (DTW\cite{vintsyuk_speech_1972, myers_comparative_1981}) based algorithm to identify the same lines in all other pixels. DTW techniques are typically used for speech recognition, where the time vector is warped to determine the correlation of two sound sequences with potentially different speeds.  Aligning the energy scale of detectors is a fundamentally very similar problem, with the filtered pulse height space of each detector getting warped to match that of a reference detector. We use a Python implementation of the FastDTW\cite{salvador_toward_2007} algorithm. Once this initial calibration is done and line positions across the broad spectrum are roughly known, a more precise determination of the line centers (and anchor points) can be done by fitting the spectral features with their known line shapes, if available. Using these anchor points we create a ``gain'' function $G$ using a cubic spline. The energy $E$ is calculated for each pulse as $E(PH) = G(PH) \times PH$, where $PH$ is the filtered pulse height. The end result is a list of time-tagged, energy-calibrated x-ray events for each individual detector. From here, the desired output product, such as an energy spectrum with particular bin widths, can easily be created.

\section{Results and Analysis}
\label{sec:Results}

\subsection{Gas Spectra}

\begin{figure*}
\includegraphics[width=1.0\linewidth]{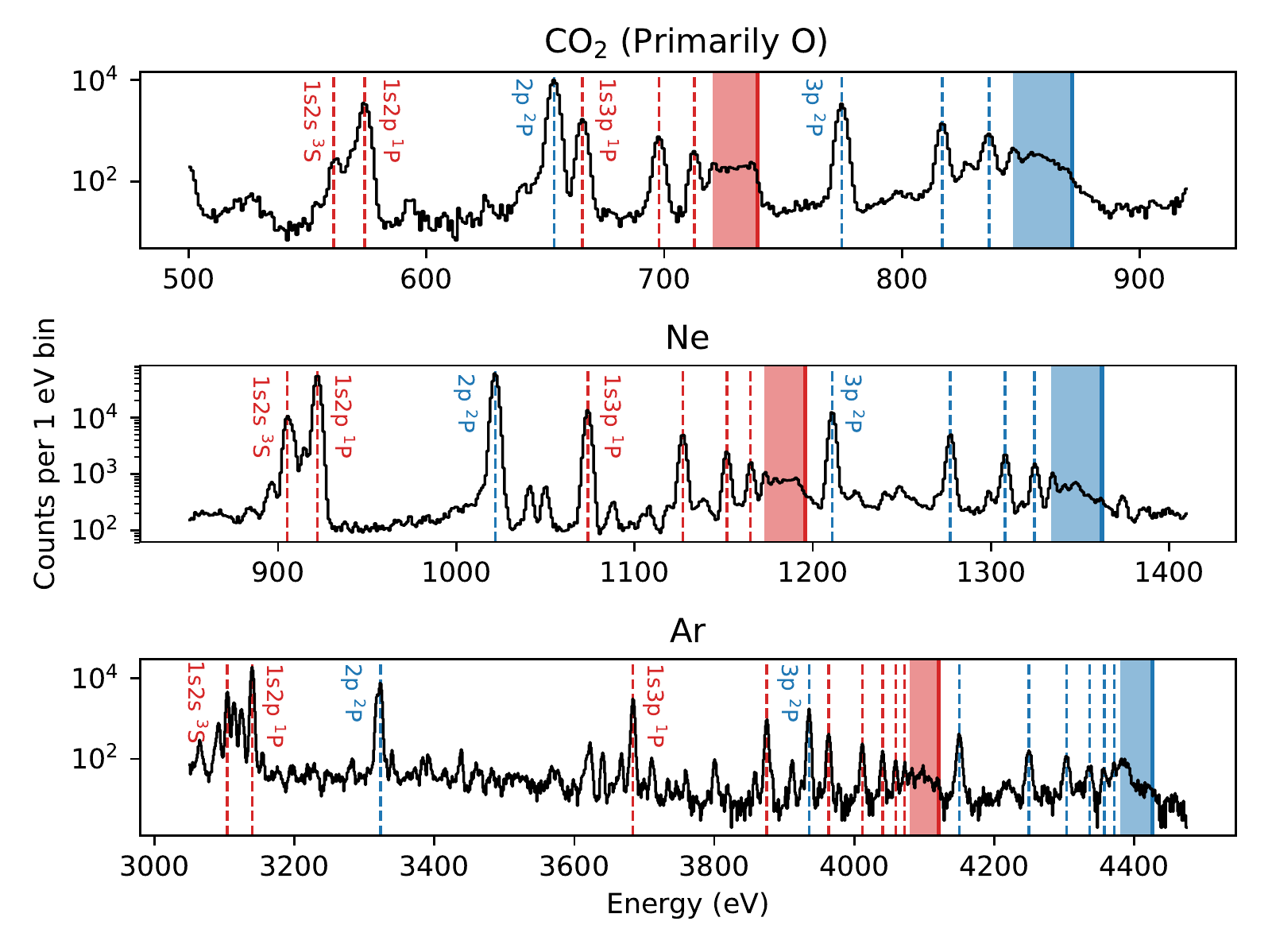}
\caption{\label{fig:CoaddedGas} Coadded CO$_2$ (primarily O), Ne, and Ar spectra with the gases ionized to primarily He-like (red) and H-like (blue) states. The left-most red dashed line corresponds to the transition from the 1s2s $^3$S excited state in He-like ions. The remaining red dashed lines make up a series of transitions from the 1s$n$p $^1$P states, where $n$ corresponds to the principal quantum number. The blue dashed lines represent 1s--$n$p transitions in H-like ions, where $n$ again corresponds to the principal quantum number. The shaded areas represent the unresolved features that make up the high energy end of the He-like and H-like 1s$n$p and $n$p series, with the solid line representing the ionization limit. In each panel, only the first 3 He-like and first 2 H-like transitions are labeled, for clarity.}
\end{figure*}

The pulse records of He-like and H-like C, O, Ne, and Ar were processed through our data reduction pipeline as outlined above. In the data with CO$_2$ gas injection, due to ionization efficiencies and poor transmittance through the Al filters at low energies, primarily O rather than C lines were observed. With all of these gases, the four most prominent lines were used for both DTW line identification and energy calibration. These were the He-like transitions to the 1s$^2$ $^1$S ground state from the 1s2p~$^1$P and 1s3p~$^1$P excited states and the H-like transitions to the 1s~$^2$S ground state from the 2p~$^2$P and 3p~$^2$P excited states. In the CO$_2$ and Ne data sets, the H-like transition to the 1s~$^2$S state from the 4p~$^2$P state was also sufficiently bright to add a calibration anchor at this transition energy. He-like and H-like line positions were located using the NIST Atomic Spectra Database (ASD)\cite{kramida_nist_nodate}, which includes a compilation of best known line positions for these ions from theory\cite{erickson_energy_1977, saloman_energy_2010} and experiment\cite{tyren_precision_1940, gabriel_interpretation_1969, peacock_spectra_1969}.

A final cut was used to remove any detectors with an outlier resolution (here, $>$5~eV measured at the H-like 2p~$^2$P line energy), as well as any detectors that failed a fit during energy calibration. Detectors that passed this final cut had their x-ray events coadded. X-ray events from the data sets of a given ion, taken across all three days of measurement, were added together. The coadded spectra of our CO$_2$, Ne, and Ar measurements are shown in Fig.~\ref{fig:CoaddedGas}. An average of 132, 141, and 125 detectors passed all cuts for the CO$_2$, Ne, and Ar data, respectively, and are coadded together in the spectra. In comparing these spectra we note that the ionization fractions are not necessarily the same, as the O spectrum contains more intense H-like lines whereas the Ar spectrum contains more intense He-like features. Nevertheless, all three spectra contain two series of lines: the He-like transitions, shown in red, and the H-like transitions, shown in blue. These relatively simple and well known series of lines are useful for testing spectrometer performance and can also be used for energy calibration purposes in future measurements.

\begin{figure}
\includegraphics[width=1.0\linewidth]{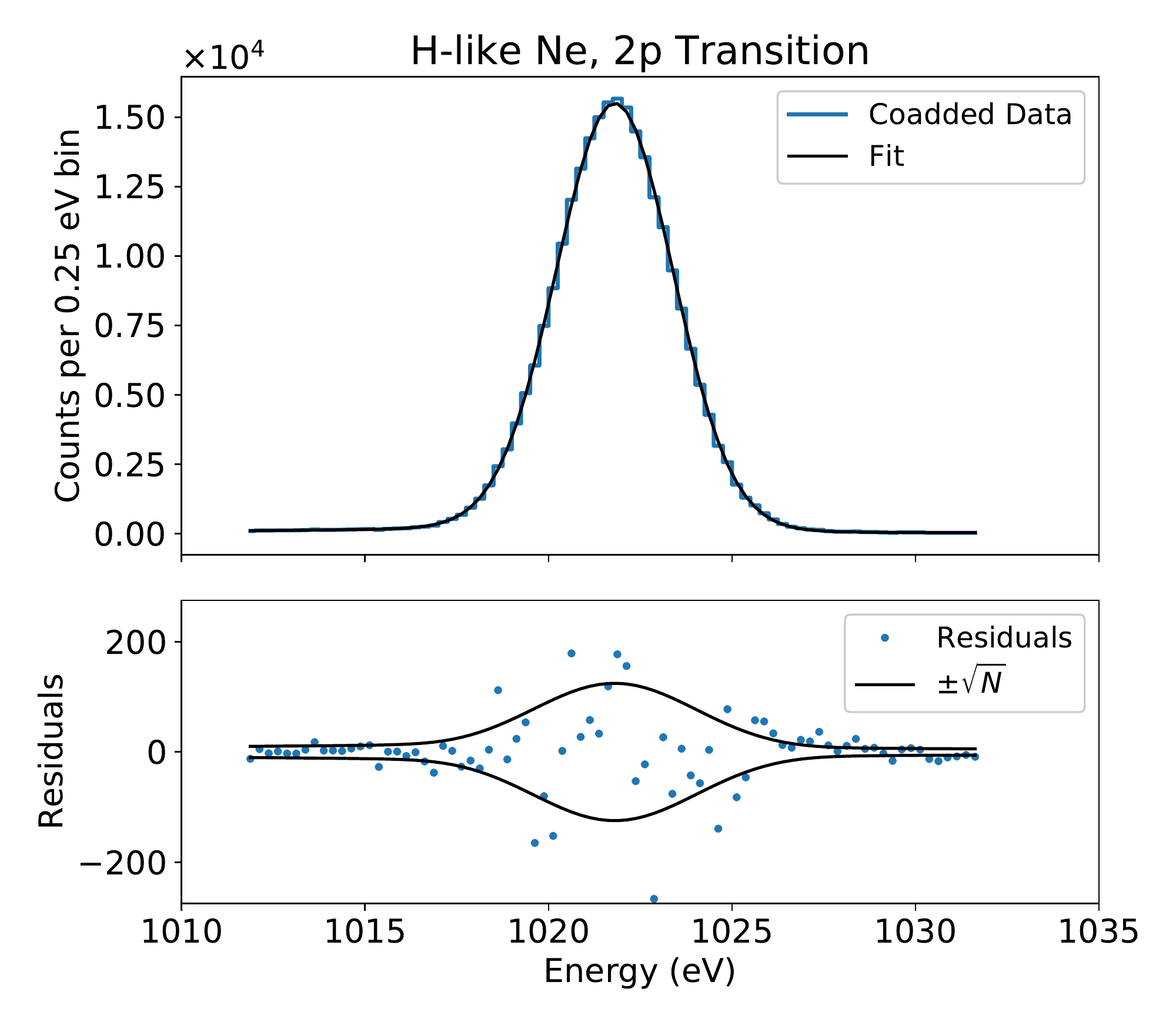}
\caption{\label{fig:FitExample} Example fit of coadded Ne data at the H-like transitions from the 2p excited state to the 1s ground state. This fit includes two narrow Lorentzians (one per transition): one from the $J = 1/2$ level at 1021.498~eV and the other from the $J = 3/2$ level at 1021.953~eV. The Lorentzians were convolved with a variable width Gaussian matching the detector response.  For this spectral feature, the fit Gaussian FWHM is $3.65 \pm 0.01$~eV. The residuals of the fit are plotted over a $\sqrt{N}$ error model characteristic of single photon counting statistics.}
\end{figure}

\begin{table*}
\caption{\centering \label{tab:GasResolutions} Gas Measurement Coadded Spectra Resolutions}
\begin{tabular}{|>{\centering\arraybackslash}m{10mm}|>{\centering\arraybackslash}m{20mm}|c|c|c|c|}
\hline
\multicolumn{2}{|c|}{Ion Species} & He-like & H-like & He-like & H-like \\
\multicolumn{2}{|c|}{Configuration} & 1s2p & 2p & 1s3p & 3p \\
\multicolumn{2}{|c|}{Term} & $^1$P & $^2$P & $^1$P & $^2$P \\
\multicolumn{2}{|c|}{$J$} & 1 & 3/2 & 1 & 3/2 \\
\hline
\multirow{2}{*}{O} & E (eV)\cite{erickson_energy_1977, tyren_precision_1940} & 573.95 & 653.68 & 665.62 & 774.63 \\
& $\Delta$E (eV) & $3.75 \pm 0.03$ & $3.73 \pm 0.02$ & $3.73 \pm 0.04$ & $3.76 \pm 0.03$ \\
\hline
\multirow{2}{*}{Ne} & E (eV)\cite{erickson_energy_1977, peacock_spectra_1969} & 922.02 & 1021.95 & 1073.77 & 1210.96 \\
& $\Delta$E (eV) & $3.63 \pm 0.01$ & $3.67 \pm 0.01$ & $3.66 \pm 0.01$ & $3.64 \pm 0.02$ \\
\hline
\multirow{2}{*}{Ar} & E (eV)\cite{saloman_energy_2010} & 3139.58 & 3322.99 & 3683.85 & 3935.72 \\
& $\Delta$E (eV) & $3.88 \pm 0.01$ & $3.89 \pm 0.02$ & $4.13 \pm 0.03$ & $4.16 \pm 0.05$ \\
\hline
\end{tabular}
\end{table*}

Features in the spectrum were fit with a model that is a convolution of a Lorentzian profile (transition with width dominated by natural line broadening) and a Gaussian profile (detector response). The convolved model is known as a Voigt profile. For these lines the Lorentzian component of this model generally has a much smaller width than the Gaussian component. For example, from transition probabilities in the He-like states\cite{johnson_e1_2002}, we expect the natural line width of the 1s2p~$^1$P--1s~$^2$S transition to be $\sim$14~meV in O, $\sim$37~meV in Ne, and $\sim$440~meV in Ar. The Gaussian FWHM (detector energy resolution), on the other hand, is $\sim$4~eV at these energies. For simplicity, we use a fixed 0.1~eV Lorentzian width (rough average for these systems) for all of our line models and note the exact choice of Lorentzian width has little impact on estimating detector response. The Lorentzian and Gaussian widths add roughly in quadrature, and even using a 0.1~eV Lorentzian width for He-like Ar 1s2p~$^1$P -- 1s~$^2$S transition instead of 0.44~eV results in only a $\sim$20~meV difference in the convolved profile width.

Many x-rays lines generated by the EBIT emulate a monochromatic source much more closely than other x-ray lines easily generated in the laboratory (such as the K lines in Sec.~\ref{subsec:CalibrationSource}). As a result, the EBIT-generated x-rays lines with simple structure, narrow transition width, and few nearby interfering lines may prove useful for studying detector behavior. For example, lines that result in a spectrum with peak to background ratio greater than 100 can be used to look for precent scale deviations from a gaussian response function, while very narrow lines may be useful for accurately measuring energy resolutions approaching 0.5~eV in future detectors optimized for $\sim$500~eV measurements\cite{morgan_use_2019}. As an example we have used 1021.95~eV 1s--2p transition in H-like Ne to measure the low energy tail fraction\cite{yan_eliminating_2017, eckart_extended_2019} in our detector response and find it to be 3\% at this energy. 

We measured the energy resolution at four distinct lines in each spectrum. The resolution here is taken to be the FWHM of the Gaussian gets convolved with Lorentzians in the fitting process. The results are shown in Table~\ref{tab:GasResolutions}. Note that the line energies listed in the table are from previous theory and measurements and are not newly measured line positions. The uncertainties listed are fitting uncertainties in the width of the Gaussian component of the fit and are typically higher in lines with a lower number of recorded counts. The resolutions are roughly 3.7~eV in the lower energy O and Ne spectra and increase to about 4~eV in the Ar data. It is generally expected that the resolution of these TES x-ray microcalorimeters will degrade somewhat with increased photon energy\cite{ullom_review_2015, doriese_practical_2017} (see Fig.~\ref{fig:CalibrationSource}, bottom), therefore it is anomalous that the O lines have slightly worse resolution than the Ne lines. We find that the slope in pulse height over energy of the detector response is higher at low energies, as expected for better resolution at lower energies. We also find that the calibration and coadding processes are not responsible for the increased resolution at the O lines. The most plausible explanation we have explored is the existence of low level interfering features coming predominatingly from H-like N and lower charge states of O, which can be close in energy to the features of interest. For example, the structure in the region of the 1s2p $^1$P line contains not only He-like O lines, but there are also lines blended into the structure originating from doubly excited Li-like O transitions. These low level features are usually not individually resolvable in the O spectrum, and it can be difficult to accurately predict their intensities in the line fit models.

The combination of high resolving power and x-ray collection efficiency make NETS particularly valuable for resolving faint lines within a broad spectrum that often contains much stronger features. As an example of this, we measured the SNR of the highest order resolvable line (defined here as being at least two FWHMs away from next highest order line and having a SNR of at least 1) in the He-like 1s$n$p and H-like $n$p transitions, where $n$ is the principal quantum number. The SNR is taken to be the amplitude of the fit function over the uncertainty in that amplitude. With this definition of resolvability in place for the O spectrum, we could fully resolve the He-like 1s5p line with a SNR of 30.6 and the H-like 5p line with a SNR of 35.6. In the Ne spectrum, we could resolve up to the He-like 1s6p line with a SNR of 48.3 and the H-like 6p line with a SNR of 48.1. Finally, in the Ar spectrum, we could resolve the transition from the He-like 1s9p state with a SNR of 8.4 and the H-like 9p state with a SNR of 4.3.

Finally, we used the He-like 1s4p $^1$P line in O, Ne, and to some extent in Ar to demonstrate the accuracy of our energy calibration procedure over narrow spectral windows. The He-like 1s4p $^1$P line was not used as anchor point to generate the calibration curves for any of the gas ion spectra. The determined energies were 697.788~eV, 1127.070~eV, and 3875.027~eV for the O, Ne, and Ar ions, respectively. Prior results\cite{tyren_precision_1940, peacock_spectra_1969, saloman_energy_2010} report these energies at 697.795~eV, 1127.095~eV, and 3874.886~eV, making our calibration as measured at the He-like 1s4p $^1$P line off by 7~meV, 25~meV, and 141~meV, respectively. This reflects the accuracy of the calibration routine when a high density of strong calibration lines are available in a narrow energy band as is the case in the H-like and He-like O and Ne data. The line placement accuracy is poorer in the higher energy H-like and He-like Ar spectrum where the accuracy is limited by low statistics in the highest energy calibration line for individual detectors and ADR cycles. We believe, based on results with similar spectrometers\cite{tatsuno_absolute_2016}, that NETS is capable of better than 141~meV line accuracy at these higher energies, and a dedicated line placement measurement would have improved the line placement accuracy here.

\subsection{Tungsten Spectrum}

\begin{figure*}[p]
\includegraphics[width=1.0\linewidth]{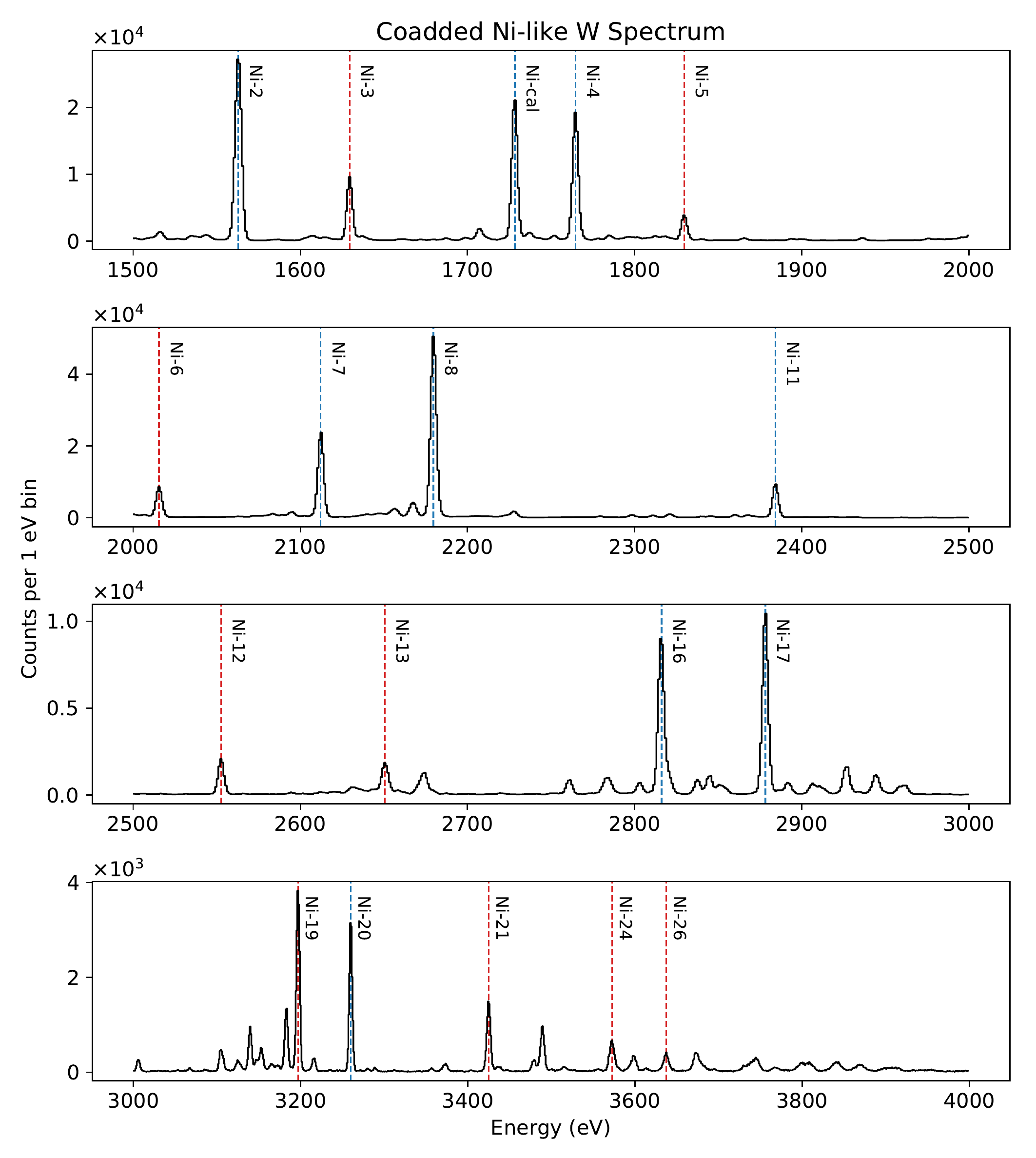}
\caption{\label{fig:CoaddedW} Ni-like W spectrum, coadded across four different data sets (total integration time of 3.46 hours) with an average of 135 detectors passing all cuts. The spectrum is separated into four regions of energy, for clarity. The blue dashed lines represent transitions that were used as anchor points in a smoothed cubic spline during the energy calibration routine. The line labeled 'Ni-cal' is the transition from the (3s$^{2}$3p$^{6}$3d$^{3}_{3/2}$3d$^{6}_{5/2}$4p$_{1/2}$)$_{J=1}$ upper level and the remaining labeled lines follow the naming convention of Clementson \textit{et al}. 2010\cite{clementson_spectroscopy_2010}. The red lines represent features whose center positions were fit as a test of the energy calibration accuracy.}
\end{figure*}

Compared to the individual gas spectra, the Ni-like W spectrum was fairly broad (1.5--4.0~keV) and required more anchor points to properly align and calibrate the entire energy range. A $\sim$2.5 keV range is typical for many of our planned future measurements.  The lower level for all reported transitions is the 3s$^{2}$3p$^{6}$3d$^{10}$ ground state. Here, we adopt the naming convention of Clementson \textit{et al}.\ 2010\cite{clementson_spectroscopy_2010} and use the Ni-2, Ni-4, Ni-11, Ni-16, Ni-17, and Ni-20 lines for detector alignment and energy calibration. In addition, we use the transition from the (3s$^{2}$3p$^{6}$3d$^{3}_{3/2}$3d$^{6}_{5/2}$4p$_{1/2}$)$_{J=1}$ level at 1728.4~eV as reported by Elliot \textit{et al}.\ 1995\cite{elliott_measurements_1995}, which was used as a calibrator in the Clementson \textit{et al}.\ 2010 work. This spans the energy range of 1562.9~eV ((3s$^{2}$3p$^{6}$3d$^{4}_{3/2}$3d$^{5}_{5/2}$4s$_{1/2}$)$_{J=3}$, or Ni-2) to 3259.9~eV ((3s$^{2}$3p$^{6}$3d$^{3}_{3/2}$3d$^{6}_{5/2}$6f$_{5/2}$)$_{J=1}$, or Ni-20).

Detectors that passed all of the data reduction cuts were coadded to form a Ni-like W spectrum. Again, detectors with resolutions worse that 5~eV are disqualified from the coadding, this time measured at 2179.7~eV ((3s$^{2}$3p$^{6}$3d$^{3}_{3/2}$3d$^{6}_{5/2}$4f$_{5/2}$)$_{J=1}$, or Ni-8). An average of 135 detectors passed all cuts for the Ni-like W analysis. The coadded spectrum is shown in Fig.~\ref{fig:CoaddedW}.

The Ni-like W measurement provides an example of the intended energy calibration routine over a wider energy range (1.5--4.0~keV) and with a more complex spectrum (many electron system) than that of the He- and H-like gas measurements. To test the calibration routine, we fit a number of spectral features with varying intensities and distances to calibration anchors. We compared the positions of these features to the same previous measurements\cite{clementson_spectroscopy_2010} that were used to fix the energy calibration anchors. Here, the goal is not to report absolute energy measurements of these transitions, but rather to check how good of a job the calibration routine can do given a set of calibration points. The results are summarized in Table~\ref{tab:LinePlacement}.

\begin{table}
\caption{\label{tab:LinePlacement} Ni-like W Energy Calibration Accuracy}
\begin{ruledtabular}
\begin{tabular}{cccc}
Upper & Measured & Ref. Exp.\cite{clementson_spectroscopy_2010} & Difference \\
Level & Energy (eV) & Energy (eV) & (eV) \\
\hline
Ni-3  & 1629.6 & 1629.8(3) & -0.2 \\
Ni-5  & 1829.7 & 1829.6(4) &  0.1 \\
Ni-6  & 2015.5 & 2015.4(4) &  0.1 \\
Ni-12 & 2552.6 & 2553.0(4) & -0.4 \\
Ni-13 & 2650.7 & 2651.3(4) & -0.6 \\
Ni-19 & 3196.8 & 3196.8(3) &  0.0 \\
Ni-21 & 3424.9 & 3426.0(4) & -1.1 \\
Ni-24 & 3572.5 & 3574.1(5) & -1.6 \\
Ni-26 & 3637.2 & 3639.5(6) & -2.3 \\
\end{tabular}
\end{ruledtabular}
\end{table}

Here, we see that the the majority of the measured energies between Ni-3 and Ni-19 fall within the measurement uncertainties reported by Clementson \textit{et al}.\ 2010\cite{clementson_spectroscopy_2010}. Lines that are far from calibration anchor points (e.g., Ni-12 and Ni-13) tended to have poorer placement accuracy than those close to anchor points (such as Ni-5 and Ni-6). In addition, the relatively low intensities of lines past 3~keV made using any lines in the region as calibration points difficult, and the Ni-20 line was the highest energy line we could use for calibration without sacrificing a majority of the detectors due to failed fits. Lines with energies above the Ni-20, such as Ni-21, Ni-24, and Ni-26, had increasingly worse placement accuracy the further in energy they were from this highest energy anchor point. This reflects on the limitations of the calibration routine for estimating the energy of pulses outside region enclosed by the anchor points where the calibration curve is unbounded. Altogether, this highlights the importance of choosing calibration targets with multiple lines close in energy to and enclosing those that are of scientific interest and will help guide us in developing calibration measurement strategies in future measurement campaigns.

\subsection{Time-resolved Analyses}

Microcalorimeters can provide valuable x-ray arrival time information that can be used to probe time-varying phenomena in an EBIT. NETS achieves its individual x-ray arrival time resolution as follows. During the initial setup of the readout electronics, a phase calibration is done to address physical and hardware latency variations and synchronize timing between readout channels to $\sim$200~ps. The readout row-to-row pixel timing uncertainty is tied to the timing jitter of the master clock, which is expected to be below a ns. These timescales are orders of magnitude below the sampling rate, which ultimately limits the x-ray arrival time resolution. NETS has a maximum sampling period of 4.8~$\mu$s when reading out all 24 rows, and this period scales linearly with the number of rows being read out. The x-ray arrival time can actually be identified with resolution exceeding the sampling period by analyzing the digitized pulse shape\cite{szymkowiak_signal_1993, fowler_practice_2016}, and a timing resolution of roughly 1~$\mu$s has been measured in a system using a nearly identical readout architecture and pulse processing pipeline\cite{heates_collaboration_first_2016}. The slew rates of the measured pulses are typically much higher than the edge trigger threshold and we do not observe integer sample delays in triggering between pixels. Generally, we do not require synchronization to an absolute timescale for our applications, but the timing can be synchronized relative to time sources of interest (such as those controlling various EBIT operations) through an external trigger signal with a precision of $\lesssim$200~ns.

\begin{figure}
\includegraphics[width=1.0\linewidth]{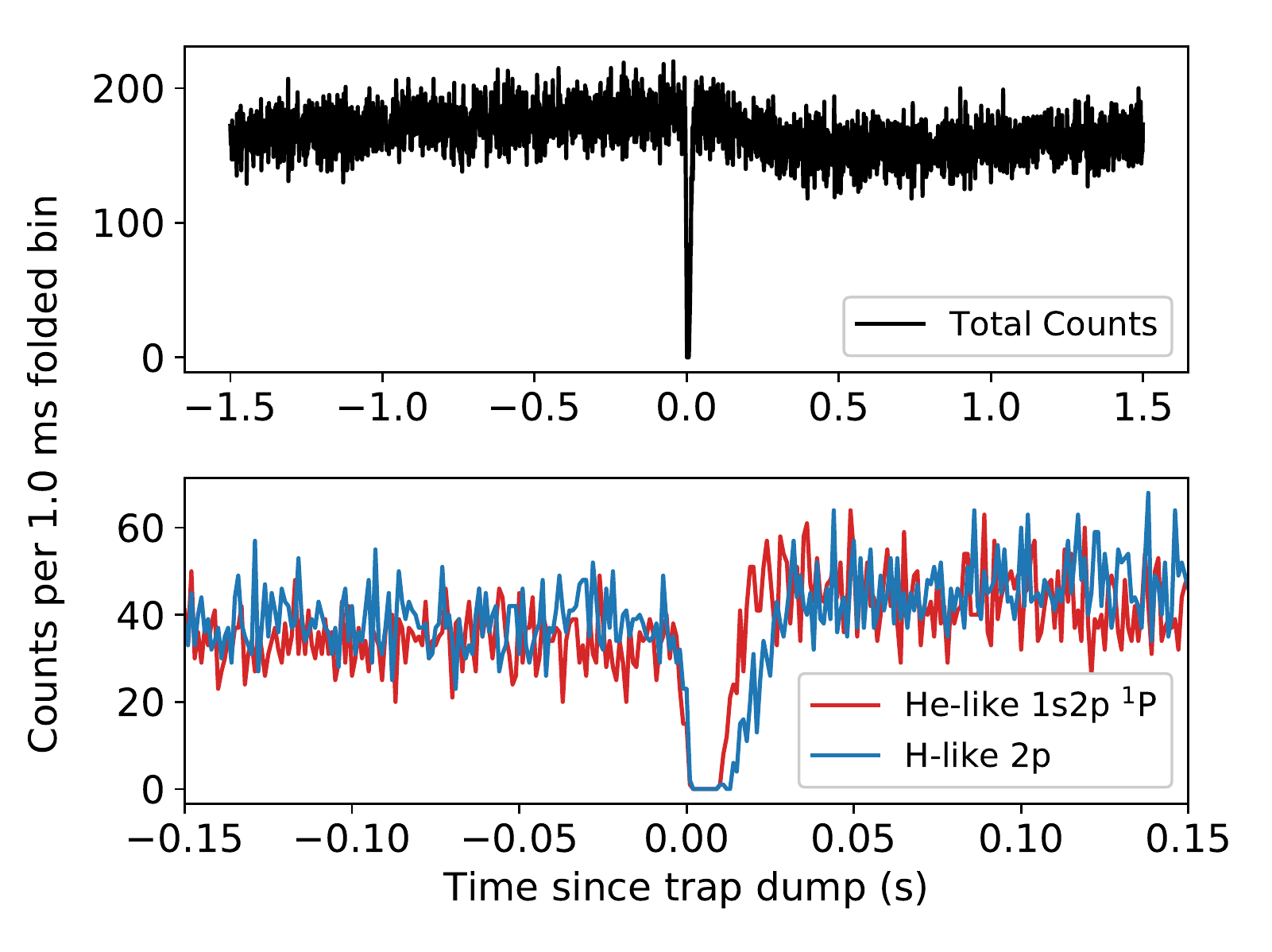}
\caption{\label{fig:TimingExample} \textit{Upper}: Total counts from a single Ne data set, folded over the 3.00002~s trap dump and gas injection period. The folded time stamps are binned into 1.0~ms bins. Here, the zero point in the time since trap dump has been set in the analysis to the point at which the falling edge crosses below 50\% of the equilibrium count rate. The width of the the dip is consistent with the 10.1~ms trap dump duration. \textit{Lower}: Counts folded over this same period separated into the He-like 1s2p $^1$P and H-like 2p states, focused on the region of time in which trap is recycled. Note that the H-like state lags behind the He-like state by $\sim$10~ms during the recovery to its equilibrium count rate.}
\end{figure}

As a simple demonstration of this capability, we took a single Ne data set and observed time-varying effects on a $\sim$3~s period due to the EBIT's trap dump and recycle process. The duration of the trap dump was set for 10.1~ms. For these early measurements, the external trigger signal had not yet been set up, so we determined the exact trap dump period using phase dispersion minimization (PDM\cite{stellingwerf_period_1978}) techniques looking for periodic drops in count rate. This accounts for uncertainty in the trap dump period as well as linear drifts between the trap dump clock and DAQ computer clock. We found the true trap dump and recycle period to be 3.00002~s and folded our measured x-ray arrival times over this period. We binned the folded time stamps into 1~ms regions and plotted the total counts versus the time since the ion trap was dumped (Fig.~\ref{fig:TimingExample}, top). The count rate drops to zero upon the trap dump, and then rises as new ions are trapped.

We also plotted the $\sim$3~s folded counts of the most prominent H-like and He-like Ne lines (Fig.~\ref{fig:TimingExample}, bottom). The goal here was to see if the two ion species evolve differently in time during a trap cycle. Although the ratio of He-like to H-like counts remained nearly constant throughout the period, the rise of the H-like state to equilibrium lagged behind the rise of the He-like state by $\sim$10~ms. This 10~ms ionization time from He-like into H-like Ne agrees very well with simple estimates of ionization rates using the well-known electron-impact cross sections at 4 keV and the typical electron density of $10^{12}$~cm$^{-3}$ for the electron beam. In future measurements, this sort of timing information can be used for determining cross sections of atomic processes and setting up optimal cycle times that maximize the counts of a desired ion state relative to other states or background.

The timing capabilities of NETS can be used for other EBIT applications where timing resolution is essential. As an example, instead of holding the beam energy of the EBIT constant, we can sweep the beam energy on roughly 10~ms timescales. As has been demonstrated by various EBIT groups, at certain beam energies and ion states, dielectronic resonances (DRs) may occur\cite{biedermann_line_2003, beilmann_high_2010, ali_photo-recombination_2011, beiersdorfer_dielectronic_2015}. The DR signal can only be produced when the beam energy is equal to the DR energy, however, it is often critical to keep the beam energy above the DR energy in order to create a desired ion population within the trap. Typically, the beam energy is held at a beam energy above the DR energy and then quickly swept through the DR energy, temporarily producing the DR signal while maintaining the desired ion population. With a time-resolving instruments such as NETS, a time folding analysis similar to the one done for Ne above can be used to isolate events at a particular timing bin, which in the case of DR measurements, could be used to isolate the DR signal. In addition to DR measurements, the time resolution can also be used to observe fundamental timescales of paired x-ray events that are correlated through certain cascading electronic transitions, among other applications.

\section{Conclusions and Future Work}
\label{sec:Conclusions}

We have successfully commissioned and took initial measurements with a TES x-ray spectrometer at the NIST EBIT. NETS improves the measurement capabilities at the NIST EBIT through a combination of high resolving power and x-ray collection efficiency. The single x-ray photon counting nature of the microcalorimeter spectrometer also provides time-resolved measurements. These capabilities will improve the accuracy and speed of ongoing measurement campaigns and in some cases enable entirely new measurements.

In order to assess the performance of NETS, we took a series of measurements of He-like and H-like O, Ne, and Ar as well as more broadband measurements of Ni-like W. We measured an energy resolution of roughly 3.7~eV at the low energy O and Ne data (0.5--1.5~keV) and about 4~eV in the higher energy Ar data (3.0--4.5~keV). With this same gas data, we were able to distinguish faint, higher order lines, up to the H-like 9p transition in Ar, in spectra containing observable features with line intensities that varied by multiple orders of magnitude. This is promising for measuring subtleties in spectral shapes and finding faint features in photon-starved measurements. We demonstrated potential energy calibration routines that could be used with NETS and found placement accuracy to better than 100~meV in narrow band spectra such as that of H-like and He-like O and Ne. With the Ni-like W data, we also demonstrated the calibration technique over a broader energy range with sparse calibration points to a few hundred meV line placement accuracy, consistent with the reference data uncertainty. Finally, we looked for periodic features with respect to the ion trap recycling time as a simple early demonstration of instrument's potential for making time-resolved measurements.

These first measurements will help to optimize data collection routines and determine which calibration measurements will be necessary for achieving a particular scientific goal. As was shown in the O and Ne data, we can calibrate line positions of unknown lines to better than 100~meV when calibration lines with well known positions are located sufficiently near the unknown lines. We can achieve this by measuring well-studied highly charged ions (such as the He-like and H-like ions) and also by making more extensive use of the external x-ray source.

In future runs, we plan on improving the capabilities of NETS in a number of ways. First, although we already see a higher x-ray collection rate compared to the NTD-Ge-based microcalorimeter that was previously attached to the NIST EBIT, this can be raised further by reducing the distance of NETS microcalorimeter array to the trap center. As can be seen in Fig.~\ref{fig:SystemIntegration}, a substantial fraction of this distance is used by the external calibration source, the size of which could be decreased fairly easily. Due to the small array diameter relative to the inner diameter of other components in the optical path, we do not expect to mask portions of the array by moving the it closer to the trap center. In terms of stray fields from the EBIT magnet, we can safely move the spectrometer $\sim$250~mm closer to the trap center before the attenuated stray field directly outside the snout starts exceeding typical background levels. In addition, using two vacuum windows between the EBIT and NETS is not strictly necessary, though it reduces some complexity in integration and risk to the various vacuum systems. Removing one of these windows could increase collection efficiency, especially at energies below $\sim$2~keV (see Fig.~\ref{fig:Efficiency}), but also by about 20\% overall due to the thick mesh component of the window. To go along with this, we plan on replacing the the 50~K filter with one that has a slightly thicker ($\sim$200~nm) Al film, but no Ni mesh, which we expect will be sufficient for transferring heat off the filter. This will have the advantage of removing a great deal of the uncertainty in the QE, but at the cost of reduced QE at the lower end of our energy band.

We also plan on improving our external calibration capabilities by introducing an annular target holder, allowing us to simultaneously collect calibration and EBIT data. This would allow us to continuously measure calibration lines, reducing the uncertainty in the drift correction associated with needing to interpolate between discrete calibration measurements separated in time. Next, we plan on improving our time-resolved measurement accuracy relative to events in the EBIT by establishing trigger signals between the two systems. Finally, we hope to increase our real-time measurement capabilities, which includes the use of the DTW alignment techniques discussed in Sec.~\ref{sec:Reduction} to produce near-final quality spectra in real-time during data acquisition. We expect these improvements to further increase the measurement speed and data quality of NETS.

\section*{Acknowledgment}
This work was supported by the NIST Innovations in Measurement Science (IMS) Program. Paul Szypryt is supported by a National Research Council Postdoctoral Fellowship. Endre Takacs is supported by the National Science Foundation Award \#1806494. The authors would like to thank Csilla Szabo-Foster, Lawrence Hudson, and Jon Pratt for their help in establishing the collaboration between the quantum sensors group and the EBIT group at NIST.


%

\end{document}